\documentclass[groupedaddress,amsmath,amssymb,notitlepage]{revtex4-1}
\pdfoutput=1

\usepackage{pdfpages}
\usepackage{etoolbox} 

\makeatletter
\patchcmd{\@outputpage@head}{\@ifx{\LS@rot\@undefined}{}{\LS@rot}}{}{}{}
\makeatother

\usepackage{graphicx}
\usepackage{dcolumn}
\usepackage{bm}
\usepackage[hidelinks]{hyperref}
\begin{document}
	


# Dual-carrier Floquet circulator with time-modulated optical resonators

Ian A. D. Williamson,[1] S. Hossein Mousavi,[1,2] Zheng Wang[1,*]

[1]Microelectronics Research Center, The University of Texas at Austin, Austin, TX 78758 USA

[2]Present address: Infinera Corporation, 140 Caspian Ct., Sunnyvale, CA 94089 USA

[*]Corresponding author: zheng.wang@austin.utexas.edu

Spatio-temporal modulation has shown great promise as a strong time-reversal symmetry breaking mechanism that enables integrated nonreciprocal devices and topological materials at optical frequencies. However, optical modulation has its own constraints in terms of modulation index and frequency, which limit the bandwidth and miniaturization of circulators and isolators, not unlike the magneto-optical schemes that it promises to replace. Here we propose and numerically demonstrate a Floquet circulator that leverages the untapped degrees of freedom unique to time-modulated resonators. Excited by sideband-selective waveguides, the system supports broadband nonreciprocal transmission without relying on the mirror or rotational symmetries required in conventional circulators. Cascading two resonators, we create a linear three-port circulator that exhibits complete and frequency-independent forward transmission between two of the ports. This approach enables wavelength-scale circulators that can rely on a variety of modulation mechanisms.

## Introduction

Nonreciprocal devices, such as circulators, play essential roles in modern optical systems to prevent feedback-induced instability in lasers [1] and amplifiers [2] and to protect interferometers [3], full-duplex transceivers [4], and read-out circuits in quantum computers from interference [5]. When combined into periodic lattices and arrays, nonreciprocal devices also provide the broken time-reversal symmetry needed to create topologically protected photonic edge states that are immune to disorder-induced backscattering [6,7]. The realization of chip-scale circulators and isolators for integrated photonics remains a grand challenge due to weak magnetic effects at optical frequencies and issues with the compatibility of magneto-optical materials and silicon photonics. Spatio-temporally modulated systems have recently emerged as a promising alternative to realize nonreciprocal responses without magnets, either via photonic transitions between spatial modes in waveguides and resonators [8], mode splitting in traveling-wave resonators [9,10], topological edge states in resonator arrays [11], or parametric modulators [12,13]. Many of these schemes have been successfully demonstrated at microwave frequencies using lumped elements [14]. However, at optical frequencies, miniaturization and broad bandwidth remain challenging for spatio-temporally modulated circulators [15] within the realistic limitations

of modulation index and available bandwidth of the mechanisms such as stimulated Brillouin scattering [16,17], cavity optomechanics [18–20], Kerr nonlinearities [21], and carrier-induced nonlinearity [22].

These challenges are fundamentally linked to the conventional structures of three-port Y-junction circulators or four-port circulators, originally conceived to harness the large frequency splitting from magnetized ferrites [23] and still retained by most spatio-temporally modulated circulators. These circulators are narrowband devices: off resonance, both forward transmission and isolation degrade rapidly and the operational bandwidth is proportional to the mode splitting. At optical frequencies, relative to the operating frequency, mode splitting from either magneto-optical effects or spatio-temporal modulation is orders of magnitude lower than that at microwave frequencies, resulting in significant bandwidth reduction, large device footprints [24], and large networks of biasing elements [7]. In addition, these circulators require high structural symmetry and fabrication order [23] to simultaneously realize complete transmission in the forward directions and large isolation.

Floquet states that arise in periodically time-modulated systems have been leveraged to break time-reversal symmetry to achieve non-reciprocity [25] and topological orders [11,26] in photonics and acoustics [7] using materials that are naturally reciprocal. In general, a time-invariant system is transformed into a Floquet system by an externally applied periodic modulation, with every static eigenstate spawning a set of Floquet states equally spaced in frequency, with relative amplitudes and phases determined by the modulating waveform. Purely sinusoidal modulation produces an amplitude distribution given by the Bessel function of the first kind, which is identical to the sideband amplitudes produced in phase modulation [27] (Fig. 1a). However, these amplitudes can be individually tailored with more general forms of modulation beyond pure single-frequency sinusoidal waveforms [28,29] (Fig. 1b). This opens up new opportunities for creating unconventional nonreciprocal responses that are not limited by the spatial symmetry or narrowband responses discussed earlier. These sideband amplitudes are well understood in the framework of parametric resonances in dynamical systems [30] and quantum field theory [31], but the deliberate control over them remains to be exploited as a new degree of freedom in nonreciprocal photonics or acoustics.

In this article, we present a compact three-port circulator based on individually tailored Floquet sidebands that provides broadband nonreciprocal transmission distinct from that of conventional circulators. We first discuss the nonreciprocal phase shift and general scattering properties of a single parametrically modulated resonator side-coupled to two waveguides. We then apply temporal coupled mode theory to study the necessary conditions that produce unique circulator responses from a low-symmetry cascade of two such Floquet resonators. Finally, a photonic crystal realization of such a Floquet circulator is presented from the results of first-principle numerical simulations.

## Results

### Floquet Resonator Nonreciprocal Phase Shifter

The building block of the proposed circulator is a single time-modulated resonator supporting a set of Floquet sidebands, which exhibits a highly nonreciprocal phase response when coupled to two narrowband waveguides. As shown in Fig. 1a, an infinite number of sidebands, i.e. Floquet states, reside in the resonator. The sidebands are spectrally distributed on both sides of the intrinsic structural resonance frequency $\omega_a$ at intervals given by the fundamental modulation frequency $\Omega$. We denote the complex instantaneous amplitude of the $n^{\text{th}}$ sideband by $a^{(n)}$. In a high quality factor resonator, where the modulation driven energy exchange between the sidebands exceeds the external coupling, the amplitudes of the sidebands are coherently correlated. Thus, the amplitude $a^{(n)}$ can be expressed as the product $a^{(n)} = u^{(n)}a$, between the *total* mode amplitude $a$ and the *relative* sideband amplitude $u^{(n)}$. $|a|^2$ is proportional to the total energy aggregated over all Floquet sidebands and $u^{(n)}$ is the solution of Hill's differential equation for the case of a general periodic modulation waveform [29,32].

Although the amplitudes of the sidebands are correlated, each sideband can couple very differently to the external environment, such that the non-trivial phase between the modulating waveform and the optical carrier waves produces a nonreciprocal response. For example, similar to the well-known channel add-drop filter configuration [33], we side-couple two parallel narrowband waveguides to either side of the Floquet resonator (Fig. 1c), with each waveguide having a narrow enough bandwidth that allows it to couple only to one of the sidebands. In the system considered here, the bottom waveguide targets the zero order sideband and the top waveguide targets the first order sideband. The evanescent coupling from the resonator to the bottom (top) waveguide results in a decay rate $\gamma_0$ $(\gamma_1)$. In a high quality factor resonator, temporal coupled mode theory (CMT) accurately describes the time evolution of the Floquet sideband amplitudes. For the system considered in Fig. 1c, but with only a single modulated resonator, two coupled mode equations are used to capture the time evolution of each sideband,

$$\frac{d}{dt}u^{(0)}a = \left(j\omega_a - \gamma\right)u^{(0)}a + \begin{pmatrix}\kappa_0 & \kappa_0\end{pmatrix}\begin{pmatrix}s_{1+} \\ s_{2+}\end{pmatrix}$$

$$\frac{d}{dt}u^{(1)}a = \left(j\omega_a + j\Omega - \gamma\right)u^{(1)}a + \begin{pmatrix}\kappa_1 & \kappa_1\end{pmatrix}\begin{pmatrix}s_{3+} \\ s_{4+}\end{pmatrix}$$

(1)

where $s_{m+}$ is the instantaneous amplitude of the incident wave from the $m^{\text{th}}$ port and $\gamma = \gamma_0 + \gamma_1 + \gamma_L$ for the loss rate due to absorption and radiation given by $\gamma_L$. Note that only two sidebands need to be explicitly

considered, and all other sidebands are uncoupled from the external environment and maintained at relative amplitudes dictated by the modulation waveform.

Unlike conventional nonreciprocal systems involving Floquet states [7], the system considered here is unique in that the signal can be transferred between multiple sidebands. To clearly differentiate the signal wave from the sideband carriers, we decompose the instantaneous amplitudes of the incoming and outgoing waves at the $m^{\text{th}}$ port, $s_{m\pm}(\omega) = \tilde{s}_{m\pm}(\Delta) e^{j(\omega_a + n\Omega)t}$ to a slowly varying envelope $\tilde{s}_{m\pm}(\Delta)$, i.e. the signal wave, and the sideband carriers $e^{j(\omega_a + n\Omega)t}$. The instantaneous frequency $\omega$ is related to the frequency detuning parameter by $\Delta = \omega - (n\Omega + \omega_a)$ where the integer $n$ is the sideband order targeted by the particular waveguide or port. The scattering parameters throughout the remainder of this paper are defined in terms of the signal wave, i.e. $S_{mp} = \tilde{s}_{m-}(\Delta) / \tilde{s}_{p+}(\Delta)$, where the optical carrier frequencies at ports $m$ and $p$ are generally different. The slowly varying envelope of the resonator $\tilde{a}$ is defined similarly.

The coefficients $\kappa_i$ are determined from time-reversal symmetry and energy conservation [34] (see Supplement). Eqn. 1 can be converted to a single expression for the transfer function between the waves and the resonator as

$$\tilde{a} = \frac{1}{j\Delta + \gamma} \begin{pmatrix} d_0 u^{(0)*} & d_0 u^{(0)*} & d_1 u^{(1)*} & d_1 u^{(1)*} \end{pmatrix} \cdot \begin{pmatrix} \tilde{s}_{1+} \\ \tilde{s}_{2+} \\ \tilde{s}_{3+} \\ \tilde{s}_{4+} \end{pmatrix}, \tag{2}$$

where $d_0$ $(d_1)$ represent the structural coupling between the bottom (top) waveguide. The associated decay rates are $\gamma_0 = |d_0 \cdot u^{(0)}|^2$ and $\gamma_1 = |d_1 \cdot u^{(1)}|^2$, which is the product of a structural coupling factor and the relative sideband amplitude (see Supplement). When absorption or radiation loss is negligible $(\gamma_L << \gamma_0, \gamma_1)$, an ideal nonreciprocal response occurs at critical coupling $(\gamma_0 = \gamma_1)$, which is equivalent to the condition

$$\left| \frac{d_0}{d_1} \right| = \left| \frac{u^{(1)}}{u^{(0)}} \right|. \tag{3}$$

An important consequence of Eqn. 3 is that any difference in the relative amplitudes of the two sidebands can be compensated by structurally asymmetric coupling (quantified by $|d_0| / |d_1|$) to achieve the ideal on-resonance response. Combining Eqn. 2 and an output coupling equation that describes the amplitudes of the outgoing

waves (see Supplement), leads to a scattering matrix for the single Floquet resonator system [34] as a function of the detuning,

$$S^I(\Delta) = \begin{pmatrix} 0 & 1 & 0 & 0 \\ 1 & 0 & 0 & 0 \\ 0 & 0 & 0 & 1 \\ 0 & 0 & 1 & 0 \end{pmatrix} - \frac{1}{j\Delta + \gamma} \begin{pmatrix} \gamma_0 & \gamma_0 & \sqrt{\gamma_0\gamma_1}e^{-j\phi} & \sqrt{\gamma_0\gamma_1}e^{-j\phi} \\ \gamma_0 & \gamma_0 & \sqrt{\gamma_0\gamma_1}e^{-j\phi} & \sqrt{\gamma_0\gamma_1}e^{-j\phi} \\ \sqrt{\gamma_0\gamma_1}e^{j\phi} & \sqrt{\gamma_0\gamma_1}e^{j\phi} & \gamma_1 & \gamma_1 \\ \sqrt{\gamma_0\gamma_1}e^{j\phi} & \sqrt{\gamma_0\gamma_1}e^{j\phi} & \gamma_1 & \gamma_1 \end{pmatrix} \qquad (4)$$

Note that the superscript I in $S^I$ denotes the case of a single Floquet resonator. The modulation phase parameter $\phi \equiv \angle u^{(1)}$ is key to a nonreciprocal response: aside from the case of $\phi$ being an integer multiple of $\pi$, $S^I$ is not symmetric. The forward pathways from the top waveguide to the bottom waveguide ($S^I_{13}$, $S^I_{14}$, $S^I_{23}$, and $S^I_{24}$) have phase shifts with opposite signs from their backward counterparts ($S^I_{31}$, $S^I_{41}$, $S^I_{32}$, and $S^I_{42}$). Specifically, signal flow in these forward pathways involves energy transfer from the $n = 0$ sideband to the $n = 1$ sideband, resulting in a positive phase shift $e^{j\phi}$, whereas the backward pathways all involve energy transfer in the *opposite* direction between the sidebands, resulting in the negative phase shift $e^{-j\phi}$. The largest nonreciprocal response is achieved when $\phi = (n + 1/2) \cdot \pi$, where $n$ is an integer. Note that with a sinusoidal modulation, $\phi$ is simply the phase difference between the modulating wave and the incoming optical carrier.

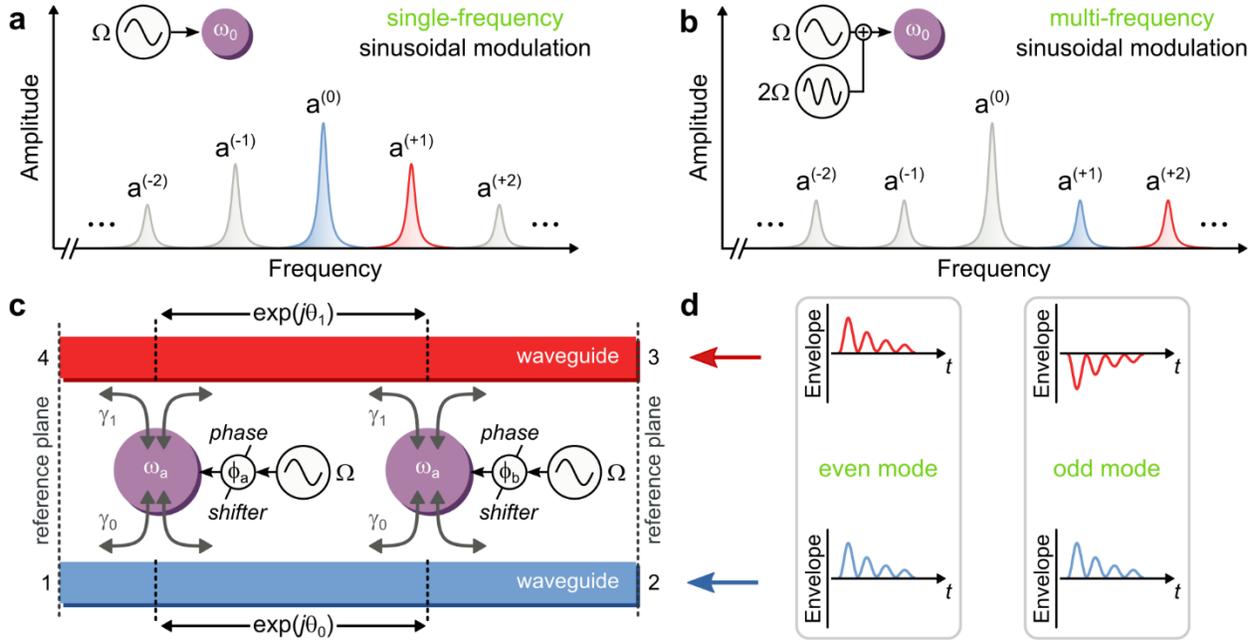

**Fig. 1.** Floquet sideband distributions and overview of the Floquet circulator system. Schematics of the Floquet sideband frequency distributions from (a) single-frequency sinusoidal modulation and (b) multi-frequency modulation waveforms. Single-frequency modulation results in a sideband amplitude distribution defined by Bessel functions of the first kind at intervals of the fundamental modulation frequency. (c) Schematic of a dual resonator Floquet circulator with resonators supporting identical modes at frequency . The resonators are modulated by single-frequency sinusoidal waveforms with fundamental frequency $\Omega$ and phases $\phi_a$ and $\phi_b$. The top waveguide (red) targets the $n = 1$ sideband with an evanescent coupling rate of $\gamma_1$, and the bottom waveguide (blue) targets the $n = 0$ sideband with an evanescent coupling rate of $\gamma_0$. The phase delay between the resonators can be different in the top and bottom waveguides, which is defined by $\theta_0$ and $\theta_1$. (d) Schematic of compound even and odd modes in the two waveguides, corresponding to a rotation of the scattering matrix into a basis defined by the vectors in Eqn. 5. The ports of the circulator are defined in compound modes on the left reference plane as $\alpha_E = \begin{pmatrix} 1 & 1 \end{pmatrix}^{\mathrm{T}}$ and $\alpha_O = \begin{pmatrix} 1 & -1 \end{pmatrix}^{\mathrm{T}}$ and on the right reference plane as $\beta_E = \begin{pmatrix} 1 & 1 \end{pmatrix}^{\mathrm{T}}$ and $\beta_O = \begin{pmatrix} 1 & -1 \end{pmatrix}^{\mathrm{T}}$, where the first (second) vector element indicates the amplitude of the wave in the bottom (top) waveguide.

## Floquet Resonator Compound Mode Filtering

To understand how the nonreciprocal *phase* response of a single resonator (Eqn. 4) can be transformed into an ideal circulator *power* response, we adopt a frame of compound waveguide modes that are linear combinations of the individual waveguide modes (Fig. 1d). We refer to the *in-phase* and *out-of-phase* combinations of the individual waveguide modes as the *even* and *odd* compound modes respectively. This notation is analogous to the Jones vector used to describe polarization states [35], in that the spatially separated individual waveguide modes at their *own* optical carrier frequencies play the role of two linear polarization components. This approach is mathematically equivalent to a rotation of the four-port scattering matrix into a basis defined by the vectors

$$\mathbf{v}_1 = \begin{pmatrix} 1 \\ 0 \\ 0 \\ 1 \end{pmatrix}, \mathbf{v}_2 = \begin{pmatrix} 1 \\ 0 \\ 0 \\ -1 \end{pmatrix}, \mathbf{v}_3 = \begin{pmatrix} 0 \\ 1 \\ 1 \\ 0 \end{pmatrix}, \mathbf{v}_4 = \begin{pmatrix} 0 \\ 1 \\ -1 \\ 0 \end{pmatrix}. \tag{5}$$

In the frame of even and odd compound modes, the single Floquet resonator acts as a notch filter, completely reflecting one polarization while completely transmitting the other. The polarization being reflected is entirely determined by the modulation phase $\phi$. For example, when the fundamental in the applied modulation waveform is in phase with the optical carrier $\left( \phi = 0 \right)$, the even compound mode $\begin{pmatrix} 1 & 1 \end{pmatrix}^T$ is completely reflected (Fig. 2a) while the odd compound mode $\begin{pmatrix} 1 & -1 \end{pmatrix}^T$ does not couple to the resonator and is completely transmitted (Fig. 2b). Tuning the applied modulation phase for maximal nonreciprocal phase response, with a quadrature phase shift with respect to the optical carrier $\left( \phi = \pi / 2 \right)$, leads to the complete reflection (Fig. 2c) of a "circular" compound mode $\begin{pmatrix} 1 & j \end{pmatrix}^T$ and the complete transmission (Fig. 2d) of the other circular compound mode $\begin{pmatrix} 1 & -j \end{pmatrix}^T$, which also does not couple to the resonator. A cascade of two Floquet resonators, along with reciprocal mode conversion in passive waveguides, then provides circulator functionality with a strong nonreciprocal response in the *scattered power*.

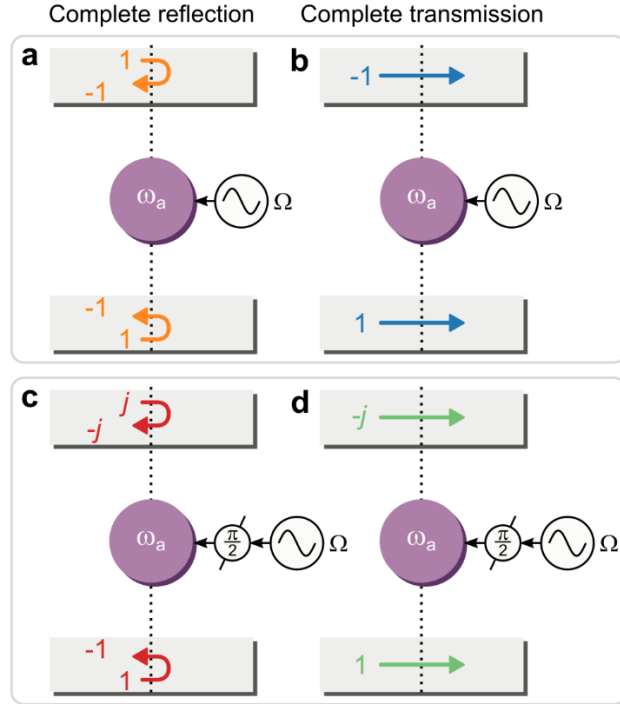

**Fig. 2.** Complete on-resonance reflection and transmission criteria for compound modes in single Floquet resonator. A modulation phase shift of $\phi = 0$ results in (a) complete reflection of the compound even mode $\begin{pmatrix} 1 & 1 \end{pmatrix}^T$ and (b) complete transmission of the compound odd mode $\begin{pmatrix} 1 & -1 \end{pmatrix}^T$. A modulation phase shift of $\phi = \pi/2$ results in (c) complete reflection of the compound even-circular mode $\begin{pmatrix} 1 & j \end{pmatrix}^T$ and (d) complete transmission of the compound odd-circular mode $\begin{pmatrix} 1 & -j \end{pmatrix}^T$.

## Dual-Resonator Floquet Circulator

The addition of a second resonator transforms the nonreciprocal response of the system to a power response. This not only eliminates the practical challenge of maintaining a locked $\pi/2$ phase shift between the modulating wave and the optical carrier, which have orders of magnitude difference in frequency, but more importantly provides a broadband circulator response useful in many applications [3,36]. The second resonator is structurally identical to the first resonator, supporting the same set of Floquet modes and is critically coupled to the same waveguides with coupling rates $\gamma_0$ and $\gamma_1$. However, the second resonator is distinct from the first resonator, with a $\pi/2$ phase lag in the modulation waveform of the second resonator, *relative to the modulation of the first.* Additionally, the bottom and top waveguides are designed to introduce different propagation phase delay between the two resonators $(\theta_0 \neq \theta_1)$. The compound modes on the left and the right reference planes constitute the four ports of the system (Fig. 1d): $\alpha_E = \begin{pmatrix} 1 & 1 \end{pmatrix}^T$ and $\alpha_O = \begin{pmatrix} 1 & -1 \end{pmatrix}^T$ on the left, $\beta_E = \begin{pmatrix} 1 & 1 \end{pmatrix}^T$ and $\beta_O = \begin{pmatrix} 1 & -1 \end{pmatrix}^T$ on the right. Here the first (second) vector element indicates the amplitude of the mode in the bottom (top) waveguide.

In the frame of the individual waveguide ports, the dual-resonator Floquet system is characterized by the following set of coupled mode equations

$$\frac{d}{dt}\begin{pmatrix} \tilde{a}_1 \\ \tilde{a}_2 \end{pmatrix} = H\begin{pmatrix} \tilde{a}_1 \\ \tilde{a}_2 \end{pmatrix} + K\begin{pmatrix} \tilde{s}_{1+} \\ \tilde{s}_{2+} \\ \tilde{s}_{3+} \\ \tilde{s}_{4+} \end{pmatrix} \tag{6}$$

$$\begin{pmatrix} \tilde{s}_{1+} \\ \tilde{s}_{2+} \\ \tilde{s}_{3+} \\ \tilde{s}_{4+} \end{pmatrix} = C\begin{pmatrix} \tilde{s}_{1+} \\ \tilde{s}_{2+} \\ \tilde{s}_{3+} \\ s_{4+} \end{pmatrix} + D\begin{pmatrix} \tilde{a}_1 \\ \tilde{a}_2 \end{pmatrix}, \tag{7}$$

where the full expressions for the coupling matrices $C$, $D$, $H$, and $K$ are given in Supplement. Here, $\tilde{a}_1$ and $\tilde{a}_2$ denote the slowly varying envelopes of the modes in the left and the right resonator respectively. The full scattering matrix as a function of the detuning frequency is [34]

$$S^{II}(\Delta) = C + D \cdot \left[ \begin{pmatrix} j\Delta & 0 \\ 0 & j\Delta \end{pmatrix} - H \right]^{-1} \cdot K. \tag{8}$$

After rotating into the compound mode basis, the on-resonance scattering matrix of the system becomes a unique circulator response

$$S^{II}_{\,c}(\Delta = 0) = \begin{pmatrix} -1 & 0 & 0 & 0 \\ 0 & 0 & 1 & 0 \\ 0 & 0 & 0 & -1 \\ 0 & 1 & 0 & 0 \end{pmatrix}. \tag{9}$$

Three of the four compound modes, ($\beta_E$, $\alpha_O$, and $\beta_O$ in Fig. 3a) form a three-port circulator. Power transmission is unity in the forward direction, defined as $\alpha_O \rightarrow \beta_O \rightarrow \beta_E \rightarrow \alpha_O$, and is completely suppressed in the backward direction. The condition for ideal operation between port $\beta_E$ and $\alpha_O$ is given by the expression $-e^{2j\theta_0}\left(e^{j\Delta\theta} - j\right)\left(e^{2j(\Delta\theta + \theta_0)} - je^{j(\Delta\theta + 2\theta_0)} + je^{j\Delta\theta} - 1\right) = 0$, where $\Delta\theta \equiv \theta_1 - \theta_0$. This is clearly satisfied by $\Delta\theta = \pi/2$, independently of the value of $\theta_0$. Ideal operation between $\alpha_O$ and $\beta_O$ translates to a slightly different expression given by $e^{2j\theta_0}\left(e^{j\Delta\theta} - j\right)\left(je^{j\Delta\theta + 2j\theta_0} + e^{2j\Delta\theta + 2j\theta_0} - je^{j\Delta\theta} - 1\right) = 0$, which is satisfied by the same condition $\Delta\theta = \pi/2$. It is also worth emphasizing that the $\alpha_E$ port is decoupled from the other three

ports, with the incoming wave being completely reflected ($\alpha_E \rightarrow \alpha_E$). See Supplement for the full derivation of the above conditions.

Conceptually, operation of the Floquet circulator can be understood by tracing a compound mode, as it travels through the system and becomes transformed. For the $\beta_E \rightarrow \alpha_O$ path (Fig. 3a, solid green curve), the signal enters as the even mode $\begin{pmatrix} 1 & 1 \end{pmatrix}^T$ on the right, and is transformed into the circular mode, $\begin{pmatrix} 1 & -j \end{pmatrix}^T$, by the dual-waveguide segment between the resonator and the ports on the right. This circular mode, $\begin{pmatrix} 1 & -j \end{pmatrix}^T$, perfectly transmits through the right resonator (as shown in Fig. 2d with $\phi = \pi/2$). Propagation down the middle waveguide segment transforms the signal into the odd mode, $\begin{pmatrix} 1 & -1 \end{pmatrix}^T$ which then impinges on the left resonator and is perfectly transmitted out the other side (as shown in Fig. 2b with $\phi = 0$). Note that the complete forward transmission between these two ports is due to the lack of coupling to either resonator, which holds even when the input signal is off-resonance. As a result, the forward transmission is unitary regardless of the detuning. The backward transmission is resonantly suppressed, with a 30 dB isolation bandwidth of $\sim 0.07\gamma$. Such broadband complete transmission in the forward direction is spectrally distinct from that of conventional three-port junction circulators and four-port circulators.

Transmission along the $\alpha_O \rightarrow \beta_O$ path (Fig. 3a, solid blue curve) can be analyzed similarly, with multiple reflections between the two resonators before the signal emerges (see Supplement). This pair of ports provides a $70\times$ larger isolation bandwidth $\left( \sim 5\gamma \right)$ than the previous pair (Fig. 3b), but at a cost of reduced forward transmission off-resonance. Backward transmission along the $\beta_O \rightarrow \alpha_O$ path (Fig. 3a, dashed blue curve) is completely suppressed on-resonance $\Delta = 0$ where the waveguides provide the ideal phase delay difference of $\Delta\theta = \pi/2$. In the off-resonance case, the difference in group velocity dispersion between the top and bottom waveguides results in incomplete mode conversion. Ultimately, $\gamma$ and the associated operating bandwidth is limited by the fundamental frequency of the modulating wave and the propagation of the coupling waveguides. The remaining ports in the system do not participate in any cross-device scattering pathways: the even mode on the left resonantly reflects into itself $\alpha_E \rightarrow \alpha_E$ (Fig. 3a, dashed orange curve), and the even mode on the right resonantly reflects into the odd mode on the right $\beta_O \rightarrow \beta_E$ (Fig. 3a, solid red curve).

An advantage of the Floquet circulator over conventional junction circulators is that its operating principal does not require breaking the degeneracy of two modes with rotational or other symmetries. The Floquet circulator has no inherent symmetry requirements in its modes, resulting in significantly simplified design and robustness towards fabrication non-idealities. More importantly, all sidebands of the two Floquet resonators remain

matched in frequency, and by introducing a functionally disconnected fourth port $(\alpha_E)$, the system achieves simultaneous *broadband* and *unity transmission* in the forward direction with perfect on-resonance isolation.

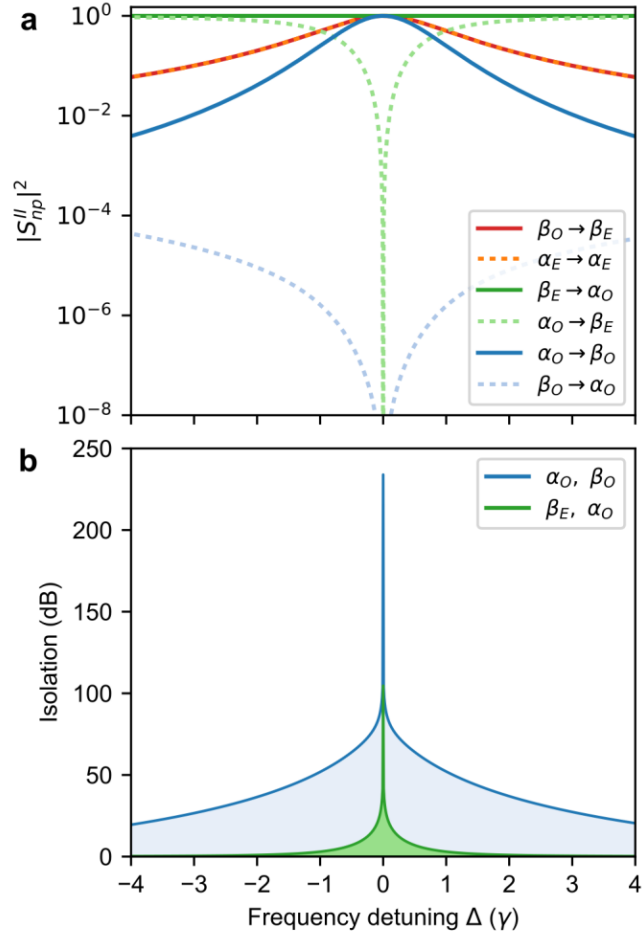

**Fig. 3.** Scattering response and isolation of an ideal Floquet circulator calculated with coupled mode theory. (a) Scattering parameter spectra for three forward signal pathways $(\beta_E \to \alpha_O \to \beta_O \to \beta_E)$, demonstrating on-resonance signal circulation. Dashed curves show suppressed backward transmission. The frequency detuning parameter $\Delta$ represents the signal frequency relative to the optical carrier, in units of $\gamma$. (b) Isolation between the two cross-device pathways $\{\alpha_O \to \beta_O, \beta_O \to \alpha_O\}$ and $\{\beta_E \to \alpha_O, \alpha_O \to \beta_E\}$.

## 2D Photonic Crystal Implementation of Floquet Circulator

We demonstrate the Floquet circulator using a 2D photonic crystal consisting of a square lattice of dielectric rods $(r = 0.15a, \epsilon_r = 12.25)$ embedded in air, which supports a large TM photonic band gap (~30 % relative bandwidth). The bandwidth of the band gap is orders of magnitude larger than the modulation rate, which eliminates radiation loss of the sidebands not targeted by the two coupling waveguides. Each parametrically modulated resonator is a point defect formed by an enlarged rod that supports a monopole mode at frequency $\omega_a = 0.39685 (2\pi c / a)$, where the refractive index of the rod is modulated by a sinusoid of frequency $\Omega = 0.0025 \omega_a$.

The two narrowband waveguides are coupled-resonator optical waveguides [37] (CROWs) formed by a 1D array of enlarged rods (Fig. 4a). The CROW period is $4a$, resulting in a propagation bandwidth of $0.0024\omega_a$, and the enlarged rod radius of each CROW is slightly different ($r_c = 0.4630a$, vs. $r_c = 0.4649a$) such that the top waveguide targets only the $n = +1$ sideband (Fig. 4b, blue) and the bottom waveguide targets only the $n = 0$ sideband (Fig. 4b, red). The $n = -1$ sideband (Fig. 4b, dashed grey) and all other sideband orders are dark to the outside environment as they couple to neither the waveguides nor the bulk crystal. The coupling rates $\gamma_0$ and $\gamma_1$ can be changed in a step-wise fashion through conventional structural tuning (i.e. adjusting the number of crystal periods between the waveguide and resonator), but as predicted by Eqn. 2 and Eqn. 3, the modulation index provides continuous tuning of $\gamma_0$ and $\gamma_1$ to realize critical coupling (Fig. 4c). Here the coupling rates are obtained from fitting the spectral response of a photonic crystal Floquet resonator to coupled mode theory. To realize the phase delay condition of $\Delta\theta = \pi/2$ in the circulator, the two resonators are separated by a distance of $16a$ (4 CROW periods).

The response of the Floquet circulator is demonstrated using first-principle finite element frequency domain simulations, with no approximations except for discretization. The electromagnetic wave equations at each sideband frequency are coupled to capture the effect of modulation and are solved simultaneously (see Supplement). As predicted by the coupled mode theory, the calculated transmission for the $\beta_E \rightarrow \alpha_O$ pathway exhibits broadband unity transmission (dark green dots, Fig. 5a) while the transmission in the backward direction $\alpha_O \rightarrow \beta_E$ is resonantly suppressed (light green dots, Fig. 5a), with over 20 dB isolation (Fig. 5b). Meanwhile, the alternative $\alpha_O \rightarrow \beta_O$ pathway exhibits unity transmission on resonance (dark blue dots, Fig. 5a) and nearly 80 dB suppression in backward transmission along $\beta_O \rightarrow \alpha_O$ (light blue dots, Fig. 5a), with a 30 dB isolation bandwidth of $\sim 4\gamma$. Note that the non-ideal line shape and reduced peak isolation observed in Fig. 5b are the result of a numerical mismatch between $\gamma_0$ and $\gamma_1$ and deviation of the waveguide phase delays from the condition $\Delta\theta = \pi/2$. Although the peak isolation is reduced to 80 dB, the broad bandwidth with above 45 dB isolation is crucial to avoiding coherence collapse in laser diodes [1]. In practice, a non-sinusoidal modulation waveform can be used to fine-tune the system closer to the critical coupling and phase delay conditions.

The strongly nonreciprocal on-resonance circulator functionality is evident in the field distributions showing the energy density aggregated over all sidebands, $\sum \left| \mathbf{E}^{(n)} \left( \Delta = 0 \right) \right|^2$. The even mode entering from port one is completely disconnected from the other ports and resonantly reflects back into itself (Fig. 5c) while the port two odd mode also resonantly reflects back but is converted into the even mode (Fig. 5d). In both cases,

transmission across the device is strongly suppressed. On the other hand, the odd mode from port one is fully transmitted into the odd mode at port two where a *resonant* scattering is observed from the greater energy in the resonator sites (Fig. 5e). As expected from the discussion earlier, the broadband forward transmission $(\beta_E \rightarrow \alpha_O)$ is clearly non-resonant, as seen in the comparable field magnitudes in the resonator and waveguide modes (Fig. 5f).

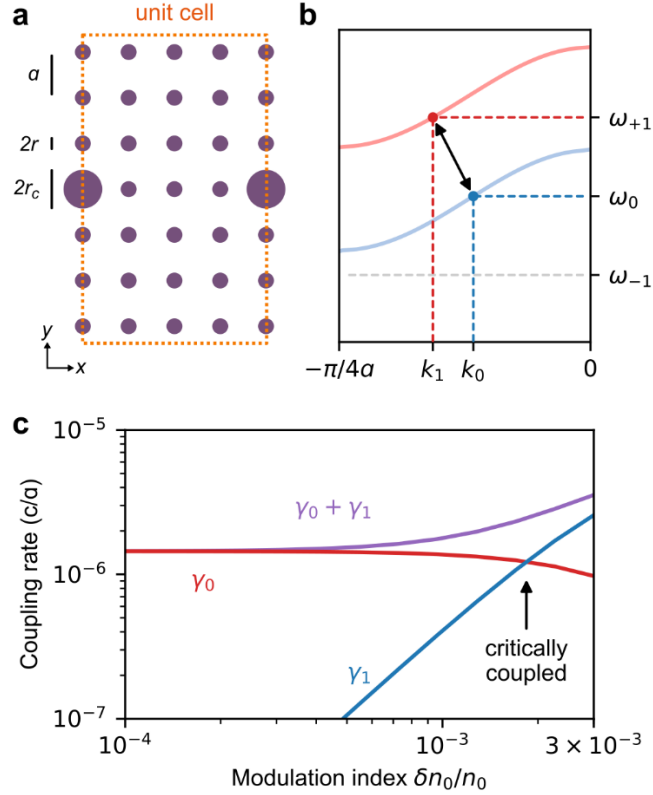

**Fig. 4.** Photonic crystal implementation of sideband-selective waveguide and critical coupling condition. (a) 2D photonic crystal consisting of silicon rods $(r = 0.15a,\ n_0 = 3.5)$ embedded in air. A coupled resonator optical waveguide (CROW) is formed in the $x$-direction by enlarging every fourth rod. (b) Dispersion for the upper (red curve) and the lower (blue curve) waveguides of Fig. 1c. The bands are detuned from each other by having $r_c = 0.4630a\ (0.4649a)$ in the top (bottom) waveguide. $\omega_n$ indicates the sideband frequency component for $n = \{-1,\ 0,\ +1\}$ and $k_n$ indicates the associated CROW wavevector. (c) Dependence of resonator coupling rate to top $(\gamma_1)$ and bottom $(\gamma_0)$ waveguide on modulation index, parameterized by the peak change in refractive index, $\delta n_0/n_0$. Critical coupling $(\gamma_0 = \gamma_1)$ occurs for $\delta n_0/n_0 = 1.84 \times 10^{-3}$. Coupling rates are calculated by fitting the response of single side-coupled modulated resonator calculated with full-wave simulations to coupled mode theory.

To operate at 1550 nm, the design's photonic crystal lattice constant is $a = 615$ nm, the modulation frequency is $\Omega = 484$ GHz, and the total quality factor is $Q = \omega_a / (2\gamma_0 + 2\gamma_1) = 8.3 \times 10^4$ (2.34 GHz linewidth). The resonator linewidth is smaller than the modulation frequency by a factor of approximately 200, meaning that the CROW bandwidth is the limiting factor for a reduced modulation frequency. Note that the simulated *in-plane* photonic crystal configuration is chosen to clearly illustrate the device operation in a compact area and a more practical modulation frequency of 20 GHz [38,39] can be used by increasing the spatial period of the CROW from $4a$ to $7a$. This reduces the propagating bandwidth to $9.3 \times 10^{-5} \omega_a$ (18 GHz), which is still nearly an order of magnitude larger than the resonator linewidth. Moreover, this results in a larger footprint in only one dimension of the system, with a total area of $56a \times 17a$ $(22.20\lambda_0 \times 6.75\lambda_0)$. Adjusting the CROW dispersion to achieve the required difference of $\pi / 2$ between the top and bottom waveguide phase delay in fewer CROW periods could further reduce the device's footprint.

The structural coupling between the resonator and the waveguides in the photonic crystal is *not* highly asymmetric, with a separation distance of $5a$ $(4a)$ between the resonators and the bottom (top) waveguide. The difference of only one unit cell requires that a relatively large modulation index be used to satisfy the critical coupling condition given by Eqn. 3, where $\delta n_0 / n_0 \approx 2 \times 10^{-3}$ leads to $u^{(1)} \approx 0.3$ (see Supplement). Although such a large modulation index is achievable in practice, for example by using the plasma dispersion effect in silicon [40], the increased absorption could overwhelm the circulator and limit the nonreciprocal response.

A more general modulating waveform that allows for independent control of the *relative* and *absolute* amplitudes of neighboring sideband states avoids the need for a large modulation index. For example, a modulating wave with a second harmonic component, as shown in Fig. 1b, can be used and the $n = 1$ and $n = 2$ sideband states can then play the roles of the $n = 0$ and $n = 1$ sideband states, respectively. The critical coupling condition can be met through the *relative ratio* between the fundamental and the second harmonic components in the modulating waveform, through the ratio $u^{(1)} / u^{(2)}$. The modulation index could then be used to achieve the necessary waveguide coupling rates that overcome out-of-plane losses and absorption (see Supplement). In contrast, single-frequency modulation with a small modulation index results in $u^{(0)} \approx 1$, and only one modulation index, satisfying $u^{(1)} \cong d_0 / d_1$, results in critical coupling for a given structure.

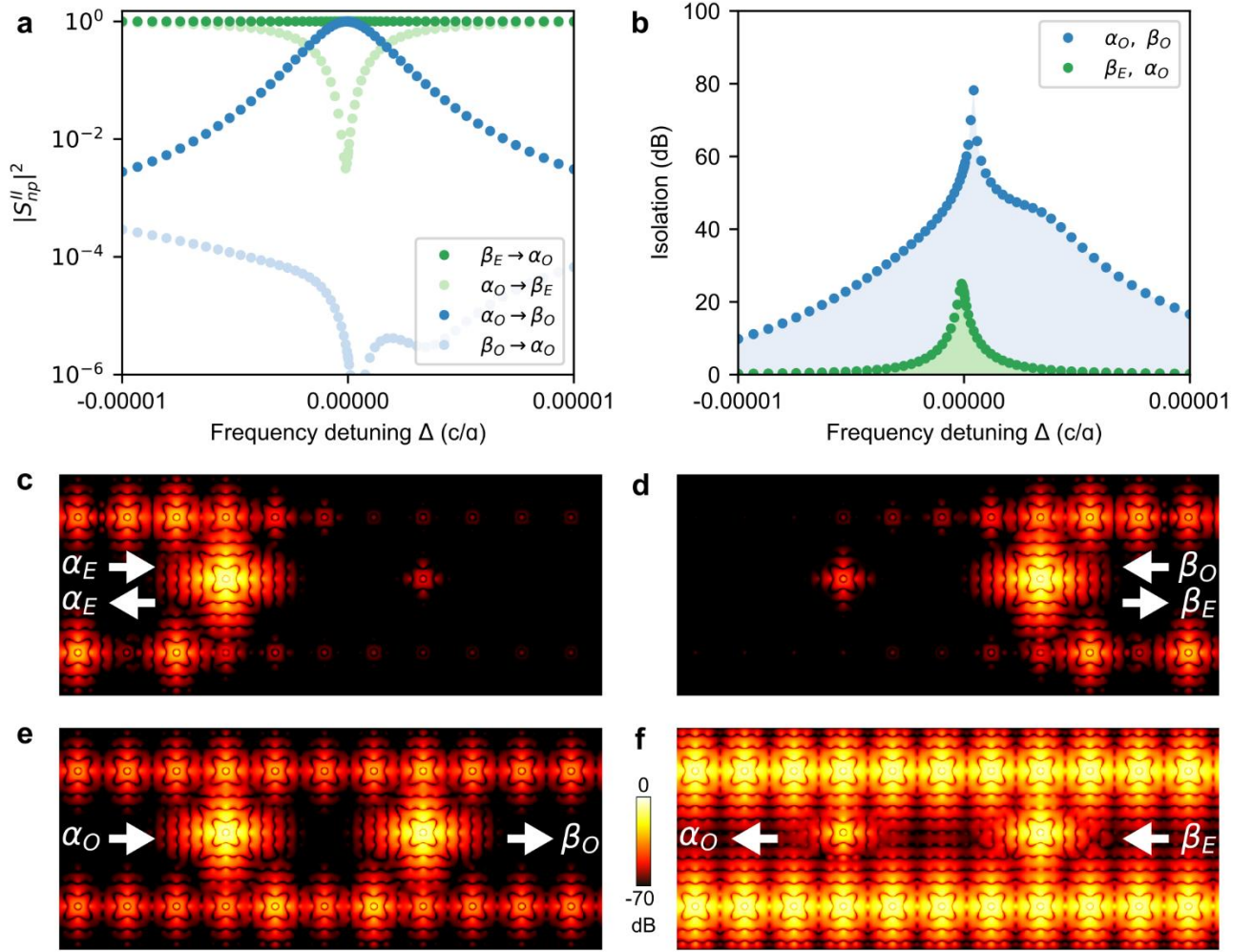

**Fig. 5.** Scattering response and field distributions of a photonic crystal Floquet circulator calculated with full-wave simulations. (a) Calculated nonreciprocal scattering parameter spectra and (b) isolation spectra for the two cross-device pathways for a Floquet circulator implemented following Fig. 4. (c-f) On-resonance electric field distributions summed over all sidebands, $\sum_n \left|\mathbf{E}^{(n)}\left(\Delta = 0\right)\right|^2$ for forward and backward incidence along the cross-device pathways. Complete reflection is evident along the backward directions in c and d, while complete transmission is observed along the forward directions in e and f.

## Conclusion

The Floquet circulator can be realized with many possible layouts of low-symmetry one-dimensional cascades of modulated resonators. The critical coupling condition and the $\Delta\theta = \pi/2$ phase shift, which collectively produce the ideal circulator response, are not tied to the rotational or mirror symmetries of the underlying photonic structures. We stress that the designation of the compound modes as *even* and *odd* is adopted from the convention in dual-waveguide systems such as directional couplers, and *does not* require that the system actually have mirror symmetry along the vertical direction. In fact, in our system the waveguides target different frequency ranges and have different coupling distances to the two resonators.

Despite a superficial structural resemblance between the Floquet circulator and circulators based on parametric up/down conversion and commutation [12,13], the signal spectral properties and overall spectral response are fundamentally different. In the Floquet circulator, the signal wave is carried by the unique dual-carrier configuration, and its spectral distribution is conserved during transit across the system (i.e. no net energy transfer occurs *between* different sideband carriers). This lack of frequency shifted intermediate signals simplifies the system design, which only needs to target a single frequency range, but also requires only two structural resonances (as opposed to four resonances with parametric conversion), resulting a smaller footprint. More importantly, without depending on large parametric conversion, the Floquet circulators can function at a modulation index that is orders of magnitude smaller than the typical values needed in parametric conversion [12,13], making it more practical for implementation at optical frequencies. Moreover, the lack of net parametric energy conversion also prevents the depletion of the modulation wave in cases involving high power signal signals. The absence of depletion ensures the system response remains linear to the input power, which is essential for high dynamic range applications [41] such as full-duplex communications [42].

The Floquet circulator can be realized with a wide range of integrated photonic platforms and modulation mechanisms. The 2D photonic crystal structure presented here can be readily extended to photonic crystal slab structures with experimentally demonstrated quality factors [43,44] exceeding $10^6$, as well as many other high-Q resonators [45]. Moreover, the two narrowband waveguides can, in principle, be merged into a single waveguide with a bandwidth spanning both the $n = 0$ and the $n = 1$ sidebands, provided that the waveguide dispersion is sufficiently large to generate the $\pi / 2$ differential phase shift. Additionally, the narrowband waveguides could also be realized with mechanisms other than a CROW, such as electromagnetically induced transparency or periodically modulated waveguides. The unusual dual-carrier response does necessitate an additional photonic circuit to modulate and demodulate traditional single-carrier optical signals to the dual-carrier format, or a spatial mode multiplexer [46] to separate the even and odd compound modes on the right-hand side of the device (see Supplement for an example). However, entire photonic circuits using such Floquet circulators can be constructed to use the dual-carrier compound modes as signal carriers [11,26], thereby limiting the need for such mode converters to the input and the output of the overall circuit.

## Funding

		
\title{Supplement}
\maketitle
\renewcommand{\theequation}{S\arabic{equation}}
\renewcommand{\thefigure}{S\arabic{figure}}
\renewcommand{\bibnumfmt}[1]{[S#1]}
\renewcommand{\citenumfont}[1]{S#1}
	
\section{Frequency-Domain Modeling of Periodically Time-Modulated Systems}
Here we outline the approach taken to model the optical Floquet system in the frequency domain. We use a set of single-frequency wave equations which are coupled in the regions that are undergoing modulation. To derive the coupling terms, we start with the general wave equation for the electric field given by
\begin{equation}
\nabla \times \nabla \times \mathbf{E}+\frac{1}{{{c}^{2}}}\frac{{{\delta }^{2}}}{\delta {{t}^{2}}}\mathbf{E}+\frac{1}{{{\epsilon }_{0}}{{c}^{2}}}\frac{{{\delta }^{2}}}{\delta {{t}^{2}}}\mathbf{P}=0,
\label{eq:wave}
\end{equation}
where $\mathbf{E}$ is the total electric field, $\mathbf{P}$ is the total electric polarizability (accounting for both static and modulated terms), $\epsilon_0$ is the vacuum permittivity, and $c$ is the speed of light in vacuum. The refractive index in the modulated region has the form $n^\prime_0=n_0+\Delta n_0 \left(t\right)$ where we assume that the time-varying component is purely sinusoidal,
\begin{equation}
\Delta {{n}_{0}}\left( t \right)=\delta {{n}_{0}}\cos \left( \Omega t+\phi  \right)=\frac{\delta {{n}_{0}}}{2}\left[ {{e}^{j\Omega t}}{{e}^{j\phi }}+{{e}^{-j\Omega t}}{{e}^{-j\phi }} \right].
\label{eq:refractive_index_time}
\end{equation}
Eqn. \ref{eq:refractive_index_time} is substituted into Eqn. \ref{eq:wave} and the wave equation is rearranged into \textit{static} and \textit{modulated} terms (where $\Delta n_0^2 \ll n_0$) to give
\begin{equation}
\nabla \times \nabla \times \mathbf{E}+\frac{{{n}_{0}}^{2}}{{{c}^{2}}}\frac{{{\delta }^{2}}}{\delta {{t}^{2}}}\mathbf{E}+\frac{1}{{{c}^{2}}}\frac{{{\delta }^{2}}}{\delta {{t}^{2}}}\left[ 2{{n}_{0}}\Delta {{n}_{0}}\left( t \right)\cdot \mathbf{E} \right]=0.
\label{eq:wave2}
\end{equation}
By representing the total electric field as a sum of components oscillating at discrete frequencies indexed by $n$,
\begin{equation}
\mathbf{E}=\sum\limits_{n}{{{\mathbf{E}}_{n}}{{e}^{j{{\omega }_{n}}t}}}
\end{equation}
and substituting into Eqn. \ref{eq:wave2}, we have
\begin{equation}
	\nabla \times \nabla \times {{\mathbf{E}}_{n}}{{e}^{j{{\omega }_{n}}t}}+\frac{{{n}_{0}}^{2}}{{{c}^{2}}}\frac{{{\delta }^{2}}}{\delta {{t}^{2}}}{{\mathbf{E}}_{n}}{{e}^{j{{\omega }_{n}}t}} 
	+ \frac{\delta {{n}_{0}}{{n}_{0}}}{{{c}^{2}}}\frac{{{\delta }^{2}}}{\delta {{t}^{2}}}\left[ \left( {{e}^{j\Omega t}}{{e}^{j\phi }}+{{e}^{-j\Omega t}}{{e}^{-j\phi }} \right){{\mathbf{E}}_{n}}{{e}^{j{{\omega }_{n}}t}} \right]=0
	\nonumber
\end{equation}
\begin{equation}
	\nabla \times \nabla \times {{\mathbf{E}}_{n}}{{e}^{j{{\omega }_{n}}t}}-\frac{{{\omega }_{n}}^{2}{{n}_{0}}^{2}}{{{c}^{2}}}{{\mathbf{E}}_{n}}{{e}^{j{{\omega }_{n}}t}} 
	- \frac{\delta {{n}_{0}}{{n}_{0}}}{{{c}^{2}}}\left[ {{\left( {{\omega }_{n}}+\Omega  \right)}^{2}}{{e}^{j\left( {{\omega }_{n+1}}t+\phi  \right)}}+{{\left( {{\omega }_{n}}-\Omega  \right)}^{2}}{{e}^{j\left( {{\omega }_{n-1}}t-\phi  \right)}} \right]{{\mathbf{E}}_{n}}=0.  
	\label{eq:wave3}
\end{equation}
The expression in Eqn. \ref{eq:wave3} is an ordinary single-frequency wave equation with an added \textit{nearest-neighbor} coupling term. Essentially, the field oscillating at frequency $\omega_n$ feeds energy into the fields oscillating at frequencies $\omega_n \pm \Omega$ through a current distribution proportional to $\mathbf{E}_n$. The coupling term is,
\begin{equation}
g\left(n,m\right)=-\frac{\delta {{n}_{0}}{{n}_{0}}}{{{c}^{2}}}{{\omega }_{n}}^{2}{{e}^{j\phi \left( n-m \right)}}\delta\left(\left|n-m\right|-1\right){{\mathbf{E}}_{m}},
\end{equation}
where the $\left(n-m\right)$ term in the exponential enforces a negative phase accumulation from a higher-order frequency component to a lower-order frequency component, and a positive phase accumulation from a lower-order frequency component to a higher-order frequency component. The Dirac delta,  $\delta\left(\left|n-m\right|-1\right)$ enforces the condition of the entire term being non-zero only when $n$ and $m$ differ by 1. This means that with a single-frequency sinusoidal modulation waveform, each sideband component has only two source terms, corresponding to the nearest neighbor frequency components, i.e. $n=0$ has non-zero terms corresponding to $g\left(0,-1\right)$ and $g\left(0,1\right)$.

This coupling can be implemented in finite element analysis by converting to a weak form expression. In this case, the weak form corresponds to a simple multiplication by the test function corresponding to the unknown electric field, e.g. $\Psi_n$. The following expression can added as a weak contribution in the modulated domain(s) for each frequency component,
\begin{equation}
F_n= g\left(n,n+1\right)\cdot\Psi_n+g\left(n,n-1\right)\cdot\Psi_n.
\end{equation}

\section{Coupled-Mode Theory (CMT) for Floquet Systems}

\begin{figure*}
	\centering
	\includegraphics{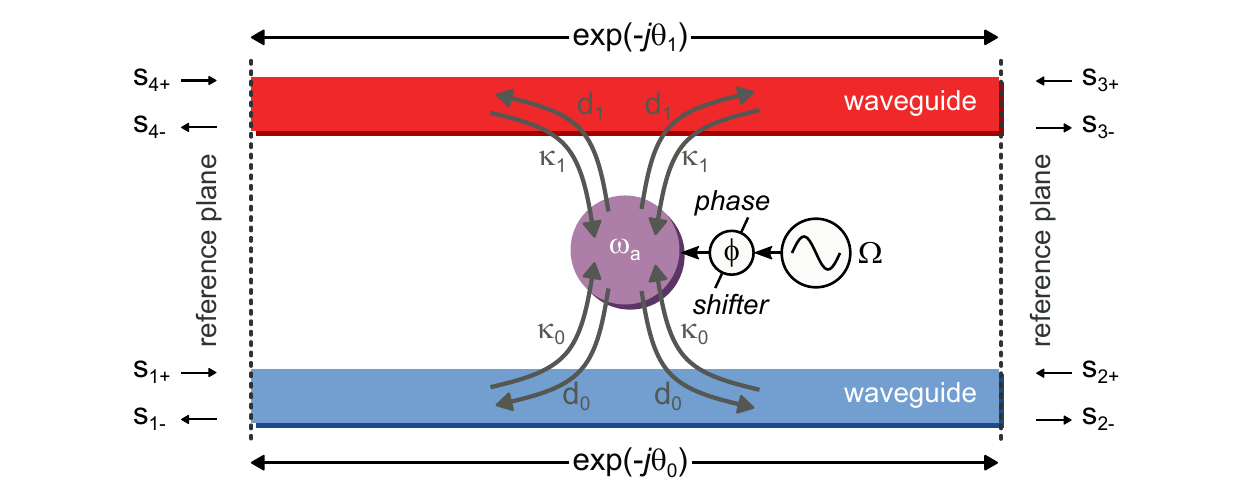}
	\caption{Schematic of single Floquet resonator with fundamental frequency $\omega_a$ modulated at a frequency of $\Omega$ and side-coupled to two narrowband waveguides. The coefficients $\kappa_i$ and $d_i$ represent the structural coupling \textit{into} and \textit{out of} the resonator, respectively. The parameters $\theta_i$ represent the phase delay of the waveguides between the port reference planes. The parameter $\phi$ represents the phase delay is applied to the modulating wave. The complex amplitudes of the incoming and outgoing waves at port $n$ are given by $s_{n\pm}$.}
	\label{fig:figS1}
\end{figure*}

The coupled-mode theory (CMT) modeling begins by considering only the amplitudes of the sidebands that couple to the external environment, in this case $n=0$ and $n=+1$.
The side-coupled resonator system is sketched in Fig. \ref{fig:figS1} and the sidebands coexist within the same resonator site. The waveguides are configured such that the top waveguide (red) is selective of the $n=+1$ sideband and the bottom waveguide (blue) is selective of the $n=0$ sideband. The two coupled-mode equations that describe the evolution in time of the sideband amplitudes are,
\begin{align}
& \frac{d}{dt}{{u}^{\left( 0 \right)}}a=
\left( j{{\omega }_{a}}-\gamma  \right){{u}^{\left( 0 \right)}}a+
\begin{pmatrix}
{{\kappa }_{0}} & {{\kappa }_{0}}  \\
\end{pmatrix}
\begin{pmatrix}
{{s}_{1+}}  \\
{{s}_{2+}}  \\
\end{pmatrix} \label{eq:cmt_in1} \\ 
& \frac{d}{dt}{{u}^{\left( 1 \right)}}a=\left( j{{\omega }_{a}}+j\Omega -\gamma  \right){{u}^{\left( 1 \right)}}a+
\begin{pmatrix}
{{\kappa }_{1}} & {{\kappa }_{1}}  \\
\end{pmatrix}
\begin{pmatrix}
{{s}_{3+}}  \\
{{s}_{4+}}  \\
\end{pmatrix}
\label{eq:cmt_in2}
\end{align}
where $s_{m+}$ is the instantaneous amplitude of the incident wave from the $m$-th port and $\gamma = \gamma_0 + \gamma_1 +\gamma_L$ for the absorption or radiation rate given by $\gamma_L$. All other sidebands are maintained at their individual amplitudes dictated by the modulating waveform. The coefficients $\kappa_{0\left( 1 \right)}$ represent the geometric coupling between the incoming wave and the $n=0$ $\left(+1\right)$ Floquet sideband. 

The system considered here is distinct from conventional nonreciprocal systems involving Floquet states, because the signal can be carried simultaneously by multiple sidebands. To clearly differentiate the signal wave from the sideband carriers, we decompose the instantaneous amplitudes of the incoming and outgoing waves at the $m$-th port, ${{s}_{m\pm }}\left( \omega  \right)={{\tilde{s}}_{m\pm }}\left( \Delta  \right){{e}^{j\left( {{\omega }_{a}}+n\Omega  \right)t}}$ to a slowly varying envelope ${{\tilde{s}}_{m\pm }}(\Delta )$, i.e. the signal wave, and the sideband carriers ${{e}^{j\left( {{\omega }_{a}}+n\Omega  \right)t}}$. The instantaneous frequency $\omega$ is related to the frequency detuning parameter by $\Delta =\omega -\left( n\Omega +{{\omega }_{a}} \right)$ where the integer $n$ is the sideband order targeted by the particular waveguide or port. 

Under this notation, the instantaneous and baseband quantities have the time dependence, 
\begin{align}
a &\sim\exp \left( j\omega t \right) &\tilde{a}&\sim\exp \left( j\Delta t \right) \\  
{{s}_{n\pm }}&\sim\exp \left( j\omega t \right)& {{{\tilde{s}}}_{n\pm }} &\sim \exp \left( j\Delta t \right).  
\end{align}
By substituting $d/dt \to j\omega$, Eqns. \ref{eq:cmt_in1} and \ref{eq:cmt_in2} can be combined into a transfer function between the incoming wave amplitudes and the overall modal amplitude given by
\begin{equation}
\tilde{a}=\frac{1}{j\Delta +\gamma}
\begin{pmatrix}
d_0 u^{\left(0\right)*} & d_0 u^{\left(0\right)*} & d_1 u^{\left(1\right)*} & d_1 u^{\left(1\right)*}
\end{pmatrix} \cdot
\begin{pmatrix}
\tilde{s}_{1+} \\
\tilde{s}_{2+} \\
\tilde{s}_{3+} \\
\tilde{s}_{4+}
\end{pmatrix},
\label{eq:cmt_xfer_in}
\end{equation}
where ${{d}_{0}}\text{ }\left( {{d}_{1}} \right)$ represent the structural coupling between the bottom (top) waveguide. The output CMT equation between the resonator and the ports is
\begin{equation}
\left( \begin{matrix}
{{{\tilde{s}}}_{1-}}  \\
{{{\tilde{s}}}_{2-}}  \\
{{{\tilde{s}}}_{3-}}  \\
{{{\tilde{s}}}_{4-}}  \\
\end{matrix} \right)=\left( \begin{matrix}
0 & {{e}^{-j{{\theta }_{0}}}} & 0 & 0  \\
{{e}^{-j{{\theta }_{0}}}} & 0 & 0 & 0  \\
0 & 0 & 0 & {{e}^{-j{{\theta }_{1}}}}  \\
0 & 0 & {{e}^{-j{{\theta }_{1}}}} & 0  \\
\end{matrix} \right)\left( \begin{matrix}
{{{\tilde{s}}}_{1+}}  \\
{{{\tilde{s}}}_{2+}}  \\
{{{\tilde{s}}}_{3+}}  \\
{{{\tilde{s}}}_{4+}}  \\
\end{matrix} \right)+\left( \begin{matrix}
{{d}_{0}}{{u}^{\left( 0 \right)}}  \\
{{d}_{0}}{{u}^{\left( 0 \right)}}  \\
{{d}_{1}}{{u}^{\left( 1 \right)}}  \\
{{d}_{1}}{{u}^{\left( 1 \right)}}  \\
\end{matrix} \right)\tilde{a},
\label{eq:cmt_out1}
\end{equation}
where ${{\theta }_{0(1)}}={{k}_{0(1)}}L$ defines the phase shift between the port reference planes in the bottom (top) waveguide for a separation distance given by $L$. Note that we have assumed the symmetry of the resonator mode results in the same coupling coefficient for port pairs that share a waveguide (i.e. $\kappa_0$, and $d_0$ for the bottom waveguide and $\kappa_1$ and $d_1$ for the top waveguide). We next use energy conservation and time reversal symmetry to derive the relationships between the linewidth, the coupling coefficients, and the relative sideband amplitudes.

\subsection{Energy Conservation}
We first consider the case of no excitation, meaning that ${{\tilde{s}}_{m+}}=0$ for all $m.$ From Eqns. \ref{eq:cmt_in1} and \ref{eq:cmt_in2}, this means that
\begin{equation}
\frac{d}{dt}{{\left| {\tilde{a}} \right|}^{2}}=-2\gamma {{\left| {\tilde{a}} \right|}^{2}}
\label{eq:cmt_energy_1}
\end{equation}
and from Eqn. \ref{eq:cmt_out1}, we have
\begin{equation}
-\frac{d}{dt}{{\left| {\tilde{a}} \right|}^{2}}=\sum{{{\left| {{s}_{m-}} \right|}^{2}}=\left( 2{{\left| {{d}_{0}}\cdot {{u}^{\left( 0 \right)}} \right|}^{2}}+2{{\left| {{d}_{1}}\cdot {{u}^{\left( 1 \right)}} \right|}^{2}} \right)}{{\left| {\tilde{a}} \right|}^{2}}.
\label{eq:cmt_energy_2}
\end{equation}
By equating the expressions in Eqns. \ref{eq:cmt_energy_1} and \ref{eq:cmt_energy_2}, we conclude that the coupling rates for the bottom and top waveguide to the resonator are given by
\begin{align}
\gamma_0&=\left| {{d}_{0}}\cdot {{u}^{\left( 0 \right)}} \right|^2 \label{eq:cmt_gamma0} \\
\gamma_1&=\left| {{d}_{1}}\cdot {{u}^{\left( 1 \right)}} \right|^2
\label{eq:cmt_gammas}
\end{align}
respectively, where (neglecting radiation and absorption loss) the total linewidth is $\gamma ={{\gamma }_{0}}+{{\gamma }_{1}}$. When absorption or radiation loss is negligible $\left( {{\gamma }_{L}}\ll {{\gamma }_{0}}\text{, }{{\gamma }_{1}} \right),$ an ideal nonreciprocal response occurs at critical coupling $\left( {{\gamma }_{0}}={{\gamma }_{1}} \right)$, which is equivalent to the condition 
\begin{equation}
\left| \frac{{{d}_{0}}}{{{d}_{1}}} \right|=\left| \frac{{{u}^{\left( 1 \right)}}}{{{u}^{\left( 0 \right)}}} \right|.
\label{eq:nr_critical}
\end{equation}
An important consequence of Eqn. \ref{eq:nr_critical} is that any difference in the relative amplitudes of the two sidebands can be compensated by structurally asymmetric coupling (quantified by the ratio $\left| {{d}_{0}} \right|/\left| {{d}_{1}} \right|$) to achieve the ideal on-resonance response.

\subsection{Time Reversal Symmetry}
The first condition provided by time reversal symmetry is
\begin{equation}
2\gamma =2{{\kappa }_{0}}{{d}_{0}}^{*}+2{{\kappa }_{1}}{{d}_{1}}^{*},
\end{equation}
which, when taken with Eqn. \ref{eq:cmt_gamma0} and Eqn. \ref{eq:cmt_gammas}, implies that ${{\kappa }_{m}}={{d}_{m}}{{\left| {{u}^{\left( m \right)}} \right|}^{2}}.$ Additionally, the direct scattering process through the waveguides requires that
\begin{equation}
\left( \begin{matrix}
{{d}_{0}}  \\
{{d}_{0}}  \\
{{d}_{1}}  \\
{{d}_{1}}  \\
\end{matrix} \right)=-\left( \begin{matrix}
0 & {{e}^{-j{{\theta }_{0}}}} & 0 & 0  \\
{{e}^{-j{{\theta }_{0}}}} & 0 & 0 & 0  \\
0 & 0 & 0 & {{e}^{-j{{\theta }_{1}}}}  \\
0 & 0 & {{e}^{-j{{\theta }_{1}}}} & 0  \\
\end{matrix} \right)\left( \begin{matrix}
{{d}_{0}}^{*}  \\
{{d}_{0}}^{*}  \\
{{d}_{1}}^{*}  \\
{{d}_{1}}^{*}  \\
\end{matrix} \right),
\label{eq:cmt_time_reversal_1}
\end{equation}
or equivalently, ${{d}_{0}}=-{{e}^{-j{{\theta }_{0}}}}{{d}_{0}}^{*}$ and ${{d}_{1}}=-{{e}^{-j{{\theta }_{1}}}}{{d}_{1}}^{*}$. By selecting the location of the reference planes such that $\theta_0$ and $\theta_1$ are some integer multiple of $2\pi$, the expression in Eqn. \ref{eq:cmt_time_reversal_1} is satisfied by $d_0$ and $d_1$ being purely imaginary and with magnitudes that satisfy Eqns. \ref{eq:cmt_gammas}. The complete expressions are therefore
\begin{align}
{{d}_{0}}&=j\frac{\sqrt{{{\gamma }_{0}}}}{\left| {{u}^{\left( 0 \right)}} \right|} \\ {{d}_{1}}&=j\frac{\sqrt{{{\gamma }_{1}}}}{\left| {{u}^{\left( 1 \right)}} \right|} \\
{{\kappa }_{0}}&={{d}_{0}}{{\left| {{u}^{\left( 0 \right)}} \right|}^{2}}=j\sqrt{{{\gamma }_{0}}}\left| {{u}^{\left( 0 \right)}} \right| \\
{{\kappa }_{1}}&={{d}_{1}}{{\left| {{u}^{\left( 1 \right)}} \right|}^{2}}=j\sqrt{{{\gamma }_{1}}}\left| {{u}^{\left( 1 \right)}} \right|.
\end{align}

\subsection{Total Scattering Matrix}
Combining Eqn. \ref{eq:cmt_xfer_in} and the output coupling relationship given by Eqn. \ref{eq:cmt_out1}, leads to a scattering matrix for the single-Floquet resonator system as a function of the detuning, 
\begin{align}
\tilde{a} &=\frac{1}{j\Delta +\gamma } 
\begin{pmatrix}
\frac{{{\kappa }_{0}}}{{{u}^{\left( 0 \right)}}} & \frac{{{\kappa }_{0}}}{{{u}^{\left( 0 \right)}}} & \frac{{{\kappa }_{1}}}{{{u}^{\left( 1 \right)}}} & \frac{{{\kappa }_{1}}}{{{u}^{\left( 1 \right)}}}  \\
\end{pmatrix} \cdot {{\begin{pmatrix}
		{{{\tilde{s}}}_{1+}} \\ {{{\tilde{s}}}_{2+}} \\ {{{\tilde{s}}}_{3+}} \\ {{{\tilde{s}}}_{4+}}  \\
		\end{pmatrix}}}\nonumber \\ 
& =\frac{1}{j\Delta +\gamma }
\begin{pmatrix}
\frac{{{d}_{0}}{{\left| {{u}^{\left( 0 \right)}} \right|}^{2}}}{{{u}^{\left( 0 \right)}}} \\
\frac{{{d}_{0}}{{\left| {{u}^{\left( 0 \right)}} \right|}^{2}}}{{{u}^{\left( 0 \right)}}} \\
 \frac{{{d}_{1}}{{\left| {{u}^{\left( 1 \right)}} \right|}^{2}}}{{{u}^{\left( 1 \right)}}} \\
  \frac{{{d}_{1}}{{\left| {{u}^{\left( 1 \right)}} \right|}^{2}}}{{{u}^{\left( 1 \right)}}}  \\
\end{pmatrix}^\text{T} \cdot {{\begin{pmatrix}
		{{{\tilde{s}}}_{1+}} \\ {{{\tilde{s}}}_{2+}} \\ {{{\tilde{s}}}_{3+}} \\ {{{\tilde{s}}}_{4+}}  \\
		\end{pmatrix}}}\nonumber \\ 
& =\frac{1}{j\Delta +\gamma } \begin{pmatrix}
j\sqrt{{{\gamma }_{0}}}{{e}^{-j\angle {{u}^{\left( 0 \right)}}}} \\ j\sqrt{{{\gamma }_{0}}}{{e}^{-j\angle {{u}^{\left( 0 \right)}}}} \\ j\sqrt{{{\gamma }_{1}}}{{e}^{-j\angle {{u}^{\left( 1 \right)}}}} \\ j\sqrt{{{\gamma }_{1}}}{{e}^{-j\angle {{u}^{\left( 1 \right)}}}}  
\end{pmatrix}^\text{T} \cdot {{\begin{pmatrix}
		{{{\tilde{s}}}_{1+}} \\ {{{\tilde{s}}}_{2+}} \\ {{{\tilde{s}}}_{3+}} \\ {{{\tilde{s}}}_{4+}}  \\
		\end{pmatrix} }}.  
\label{eq:cmt_in_final} 
\end{align}
The output coupling is given by the relationship
\begin{align}
\begin{pmatrix}
{{{\tilde{s}}}_{1-}}  \\
{{{\tilde{s}}}_{2-}}  \\
{{{\tilde{s}}}_{3-}}  \\
{{{\tilde{s}}}_{4-}}  \\
\end{pmatrix}
&=\begin{pmatrix}
0 & {{e}^{-j{{\theta }_{0}}}} & 0 & 0  \\
{{e}^{-j{{\theta }_{0}}}} & 0 & 0 & 0  \\
0 & 0 & 0 & {{e}^{-j{{\theta }_{1}}}}  \\
0 & 0 & {{e}^{-j{{\theta }_{1}}}} & 0  \\
\end{pmatrix}
\begin{pmatrix}
{{{\tilde{s}}}_{1+}}  \\
{{{\tilde{s}}}_{2+}}  \\
{{{\tilde{s}}}_{3+}}  \\
{{{\tilde{s}}}_{4+}}  \\
\end{pmatrix} +\begin{pmatrix}
{{d}_{0}}{{u}^{\left( 0 \right)}}  \\
{{d}_{0}}{{u}^{\left( 0 \right)}}  \\
{{d}_{1}}{{u}^{\left( 1 \right)}}  \\
{{d}_{1}}{{u}^{\left( 1 \right)}}  \\
\end{pmatrix} \tilde{a} \nonumber \\ 
&=\begin{pmatrix}
0 & {{e}^{-j{{\theta }_{0}}}} & 0 & 0  \\
{{e}^{-j{{\theta }_{0}}}} & 0 & 0 & 0  \\
0 & 0 & 0 & {{e}^{-j{{\theta }_{1}}}}  \\
0 & 0 & {{e}^{-j{{\theta }_{1}}}} & 0  \\
\end{pmatrix}
\begin{pmatrix}
{{{\tilde{s}}}_{1+}}  \\
{{{\tilde{s}}}_{2+}}  \\
{{{\tilde{s}}}_{3+}}  \\
{{{\tilde{s}}}_{4+}}  \\
\end{pmatrix} +
\begin{pmatrix}
j\sqrt{{{\gamma }_{0}}}{{e}^{j\angle {{u}^{\left( 0 \right)}}}}  \\
j\sqrt{{{\gamma }_{0}}}{{e}^{j\angle {{u}^{\left( 0 \right)}}}}  \\
j\sqrt{{{\gamma }_{1}}}{{e}^{j\angle {{u}^{\left( 1 \right)}}}}  \\
j\sqrt{{{\gamma }_{1}}}{{e}^{j\angle {{u}^{\left( 1 \right)}}}}  \\
\end{pmatrix}\tilde{a}. 
\label{eq:cmt_out_final} 
\end{align}
By combining Eqn. \ref{eq:cmt_in_final}  with Eqn. \ref{eq:cmt_out_final} and letting $\phi =\angle {{u}^{\left( 1 \right)}}-\angle {{u}^{\left( 0 \right)}}$, the complete scattering matrix for the system can be solved for,
\begin{equation}
S^{\text{I}}\left( \Delta  \right)=\left( \begin{matrix}
0 & 1 & 0 & 0  \\
1 & 0 & 0 & 0  \\
0 & 0 & 0 & 1  \\
0 & 0 & 1 & 0  \\
\end{matrix} \right)-
\frac{1}{j\Delta +\gamma }\left( \begin{matrix}
{{\gamma }_{0}} & {{\gamma }_{0}} & \sqrt{{{\gamma }_{0}}{{\gamma }_{1}}}{{e}^{-j\phi }} & \sqrt{{{\gamma }_{0}}{{\gamma }_{1}}}{{e}^{-j\phi }}  \\
{{\gamma }_{0}} & {{\gamma }_{0}} & \sqrt{{{\gamma }_{0}}{{\gamma }_{1}}}{{e}^{-j\phi }} & \sqrt{{{\gamma }_{0}}{{\gamma }_{1}}}{{e}^{-j\phi }}  \\
\sqrt{{{\gamma }_{0}}{{\gamma }_{1}}}{{e}^{j\phi }} & \sqrt{{{\gamma }_{0}}{{\gamma }_{1}}}{{e}^{j\phi }} & {{\gamma }_{1}} & {{\gamma }_{1}}  \\
\sqrt{{{\gamma }_{0}}{{\gamma }_{1}}}{{e}^{j\phi }} & \sqrt{{{\gamma }_{0}}{{\gamma }_{1}}}{{e}^{j\phi }} & {{\gamma }_{1}} & {{\gamma }_{1}}  \\
\end{matrix} \right).
\label{eq:scattering_1}
\end{equation}
Note that the superscript $I$ in ${{S}^{I}}$ denotes the case of a single Floquet resonator.

\section{Dual Floquet Resonators}
In the frame of the individual waveguide ports, the dual-resonator Floquet system is characterized by the following set of coupled mode equations
\begin{equation}
\frac{d}{dt}
\begin{pmatrix}
{{{\tilde{a}}}_{1}}  \\
{{{\tilde{a}}}_{2}}  \\
\end{pmatrix} =H
\begin{pmatrix}
{{{\tilde{a}}}_{1}}  \\
{{{\tilde{a}}}_{2}}  \\
\end{pmatrix} +K
\begin{pmatrix}
{{{\tilde{s}}}_{1+}}  &
{{{\tilde{s}}}_{2+}}  &
{{{\tilde{s}}}_{3+}}  &
{{{\tilde{s}}}_{4+}}  
\end{pmatrix}^{\text{T}}
\end{equation}

\begin{equation}
\begin{pmatrix}
{{{\tilde{s}}}_{1+}}  \\
{{{\tilde{s}}}_{2+}}  \\
{{{\tilde{s}}}_{3+}}  \\
{{{\tilde{s}}}_{4+}}  \\
\end{pmatrix} =C
\begin{pmatrix}
{{{\tilde{s}}}_{1+}}  \\
{{{\tilde{s}}}_{2+}}  \\
{{{\tilde{s}}}_{3+}}  \\
{{s}_{4+}}  \\
\end{pmatrix}+D
\begin{pmatrix}
{{{\tilde{a}}}_{1}}  \\
{{{\tilde{a}}}_{2}}  \\
\end{pmatrix},
\end{equation}
where ${{\tilde{a}}_{1}}$ and ${{\tilde{a}}_{2}}$ denote the slowly varying envelopes of the modes in the left and the right resonator, respectively and $\phi_a$ and $\phi_b$ are their associated modulation phases. The self- and inter-resonator coupling matrix is 
\begin{equation}
H=\begin{pmatrix}
j\omega_a  & j\mu  \\
j\mu & j\omega_b   \\
\end{pmatrix} 
-\begin{pmatrix}
\gamma  & \gamma_0 e^{-j\theta_0}+\gamma_1 e^{-j\theta_1}e^{j(\phi_b-\phi_a}  \\
\gamma_0e^{-j\theta_0}+\gamma_1e^{-j\theta_1}e^{j(\phi_a-\phi_b)} & \gamma   \\
\end{pmatrix},
\end{equation}
the input port coupling matrix is
\begin{equation}
K=\begin{pmatrix}
j\sqrt{{{\gamma }_{0}}} & j\sqrt{{{\gamma }_{0}}}{{e}^{-j{{\theta }_{0}}}} & j\sqrt{{{\gamma }_{1}}}{{e}^{-j{{\theta }_{1}}}}{{e}^{-j{{\phi }_{a}}}} & j\sqrt{{{\gamma }_{1}}}{{e}^{-j{{\phi }_{a}}}}  \\
j\sqrt{{{\gamma }_{0}}}{{e}^{-j{{\theta }_{0}}}} & j\sqrt{{{\gamma }_{0}}} & j\sqrt{{{\gamma }_{1}}}{{e}^{-j{{\phi }_{b}}}} & j\sqrt{{{\gamma }_{1}}}{{e}^{-j{{\theta }_{1}}}}{{e}^{-j{{\phi }_{b}}}}  \\
\end{pmatrix},
\end{equation}
the output port coupling matrix is
\begin{equation}
D=\begin{pmatrix}
j\sqrt{{{\gamma }_{0}}} & j\sqrt{{{\gamma }_{0}}}{{e}^{-j{{\theta }_{0}}}}  \\
j\sqrt{{{\gamma }_{0}}}{{e}^{-j{{\theta }_{0}}}} & j\sqrt{{{\gamma }_{0}}}  \\
j\sqrt{{{\gamma }_{1}}}{{e}^{-j{{\theta }_{1}}}}{{e}^{j{{\phi }_{a}}}} & j\sqrt{{{\gamma }_{1}}}{{e}^{j{{\phi }_{b}}}}  \\
j\sqrt{{{\gamma }_{1}}}{{e}^{j{{\phi }_{a}}}} & j\sqrt{{{\gamma }_{1}}}{{e}^{-j{{\theta }_{1}}}}{{e}^{j{{\phi }_{b}}}}  \\
\end{pmatrix},
\end{equation}
and the direct port-to-port scattering matrix is
\begin{equation}
C=\begin{pmatrix}
0 & {{e}^{-j{{\theta }_{0}}}} & 0 & 0  \\
{{e}^{-j{{\theta }_{0}}}} & 0 & 0 & 0  \\
0 & 0 & 0 & {{e}^{-j{{\theta }_{1}}}}  \\
0 & 0 & {{e}^{-j{{\theta }_{1}}}} & 0  \\
\end{pmatrix}.
\end{equation}
$\theta_0$ and $\theta_1$ are the transmission phases of the bottom and top waveguide shown in Fig. \ref{fig:figS1}. Note that $D\ne {{K}^{\text{T}}}$ and $\gamma ={{\gamma }_{0}}+{{\gamma }_{1}}+{{\gamma }_{L}}$. Throughout this work we assume that the two resonators are structurally identical $\left(\omega_b=\omega_a\right)$ and that evanescent coupling between the two resonator sites is negligible $\left(\mu=0\right)$. The full scattering matrix as a function of the detuning frequency is
\begin{equation}
S^{II}\left( \Delta  \right)=C+D\cdot {{\left[ 
		\begin{pmatrix}
		j\Delta  & 0  \\
		0 & j\Delta   \\
		\end{pmatrix}-H \right]}^{-1}}\cdot K
\end{equation}
where the superscript $II$ denotes a cascade of two Floquet resonators.

\section{Analytical Expressions for Scattering Parameters}
Considering the relationship $\theta_0 = \theta_1 + \Delta\theta$, the condition for ideal isolation between $\alpha_O$ and $\beta_O$ is
\begin{equation}
{{e}^{2j{{\theta }_{\text{0}}}}}\left( {{e}^{j\Delta \theta }}-j \right)\left( j{{e}^{j\Delta \theta +2j{{\theta }_{\text{0}}}}}+{{e}^{2j\Delta \theta +2j{{\theta }_{\text{0}}}}}-j{{e}^{j\Delta \theta }}-1 \right)=0
\end{equation}
which is satisfied by $\Delta\theta = \pi/2$, independently of the value of $\theta_0$. Ideal operation between $\beta_E$ and $\alpha_O$ translates to a slightly different expression given by
\begin{equation}
-{{e}^{2j{{\theta }_{\text{0}}}}}\left( {{e}^{j\Delta \theta }}-j \right)\left( {{e}^{2j(\Delta \theta +{{\theta }_{0}})}}-j{{e}^{j(\Delta \theta +2{{\theta }_{0}})}}+j{{e}^{j\Delta \theta }}-1 \right)=0
\end{equation}
which is satisfied by the same condition $\Delta\theta = \pi/2$. The overall dispersive scattering matrix of the system becomes a circulator response given by
\begin{equation}
S^{\text{II}}_c=
\begin{pmatrix}
-\frac{\gamma }{\gamma +j \Delta } & -\frac{j \gamma  \Delta }{\left(\gamma +j \Delta \right)^2} & 0 & \frac{\Delta ^2}{\left(\Delta -j \gamma \right)^2} \\
0 & 0 & 1 & 0 \\
0 & \frac{j \Delta }{\gamma +j \Delta } & 0 & -\frac{\gamma }{\gamma +j \Delta } \\
\frac{j \Delta }{\gamma +j \Delta } & \frac{\gamma ^2}{\left(\gamma +j \Delta \right)^2} & 0 & -\frac{j \gamma  \Delta }{\left(\gamma +j \Delta \right)^2} 
\end{pmatrix}.
\label{eq:circ_scattering}
\end{equation}
The scattering matrix in Eqn. \ref{eq:circ_scattering} is distinct from the scattering response of a three-port junction circulator \cite{lax_microwave_1962}, where transmission in one of the forward direction has the form 
\begin{equation}
T=\frac{2}{3}\left( \frac{{{e}^{j4\pi /3}}}{1+j\left( \omega -{{\omega }_{a}} \right)/{{\gamma }_{a}}}+\frac{{{e}^{j2\pi /3}}}{1+j\left( \omega -{{\omega }_{b}} \right)/{{\gamma }_{b}}} \right).
\end{equation}

\section{Nonreciprocal Signal Pathways in Dual-Resonator Circulator}
\begin{figure}[ht!]
	\centering
	\includegraphics{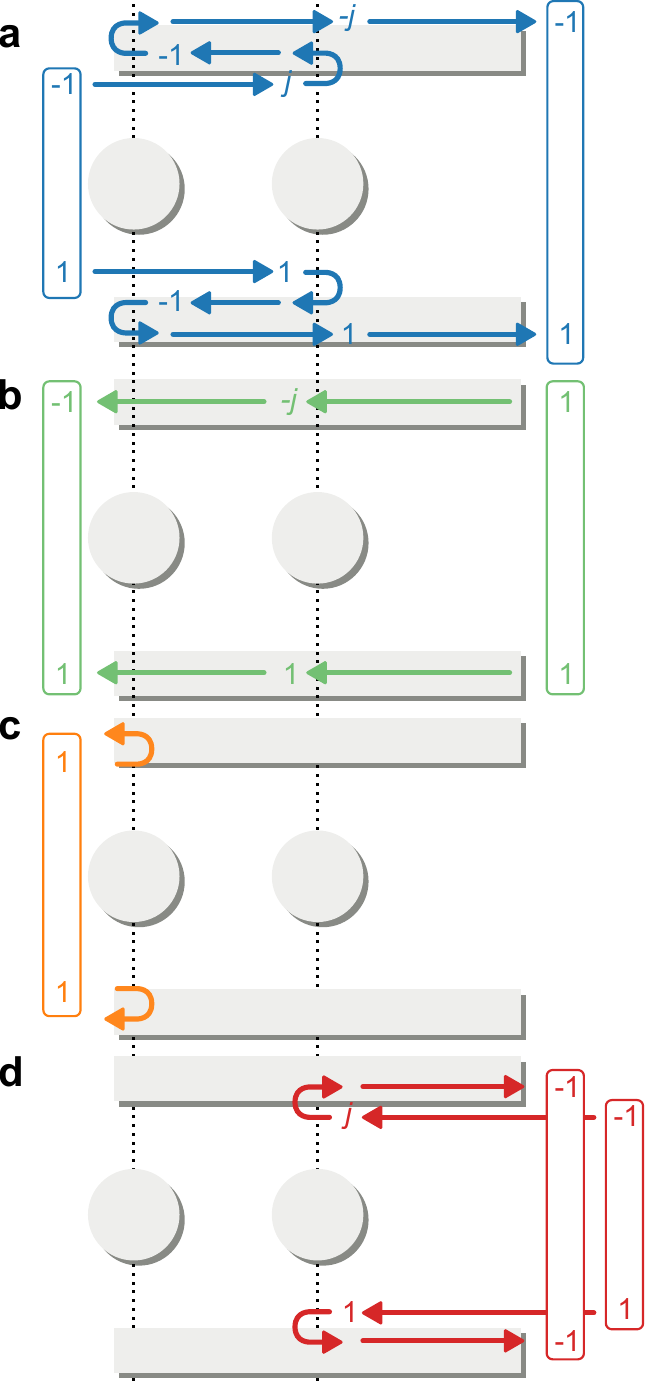}
	\caption{(a) Incidence of odd mode from the left undergoes two reflections before being resonantly transmitted out of the right side. (b) Incidence of the even mode from the right is non-resonantly transmitted to the odd mode on the left. (c) Incidence of the even mode from the left is resonantly reflected back into the even mode. (d) Incidence of the odd mode from the right is resonantly reflected back into the even mode on the right.}
	\label{fig:figS2}
\end{figure}

The on-resonance operation of the Floquet circulator can be visualized by tracing a compound mode as it travels through the system and becomes transformed by the segments of waveguides and as it reflects and transmits through the resonators (Fig. \ref{fig:figS2}). For the $\beta_E\to\alpha_O$ path, the signal enters as the even mode on the right, and is transformed into the circular mode, by the dual-waveguide segment between the resonator and the ports on the right. This circular mode, perfectly transmits through the right resonator (as shown in Fig. 2d of the main text with $\phi=\pi/2$). Propagation down the middle waveguide segment transforms the signal into the odd mode, which then impinges on the left resonator and is perfectly transmitted out the other side (as shown in Fig. 2b of the main text with $\phi=0$). Note that the complete forward transmission from $\beta_E\to\alpha_O$ is due to the lack of coupling to either resonator, which holds even when the input signal is off-resonance. As a result, the forward transmission is unitary regardless of the detuning as expected from the expression in Eqn. \ref{eq:circ_scattering}. The backward transmission is resonantly suppressed, with a 30 dB isolation bandwidth of $\sim0.07\gamma$. Such broadband complete transmission in the forward direction is spectrally distinct from that of conventional junction circulators.

Transmission along the $\alpha_O\to\beta_O$ path can be analyzed similarly, with multiple reflections between the two resonators before the signal emerges (Fig. \ref{fig:figS2}). This pair of ports provides a $70\times $ larger isolation bandwidth $\left(\sim5\gamma\right)$ than the previous pair, but at a cost of reduced forward transmission off-resonance. Ultimately, $\gamma$ and the associated operating bandwidth is limited by the fundamental frequency of the modulating wave and the and the propagation of the coupling waveguides. The remaining ports in the system do not participate in any cross-device scattering pathways: the even mode on the left resonantly reflects into itself $\alpha_E\to\alpha_E$, and the even mode on the right resonantly reflects into the odd mode on the right $\beta_O\to\beta_E$.

On-resonance this results in the response
\begin{equation}
S^{\text{II}}_c\left(\Delta=0\right)=
\begin{pmatrix}
-1 & 0 & 0 & 0  \\
0 & 0 & 1 & 0  \\
0 & 0 & 0 & -1  \\
0 & 1 & 0 & 0  \\
\end{pmatrix},
\end{equation}
while off-resonance the scattering is reciprocal with the response
\begin{equation}
S^{\text{II}}_c\left(\Delta\rightarrow\infty\right)=
\begin{pmatrix}
0 & 0 & 0 & 1  \\
0 & 0 & 1 & 0  \\
0 & 1 & 0 & 0  \\
1 & 0 & 0 & 0  \\
\end{pmatrix}.
\end{equation}

\section{Sideband Amplitude Distribution from Full-wave Simulation}
\begin{figure}[ht!]
	\centering
	\includegraphics{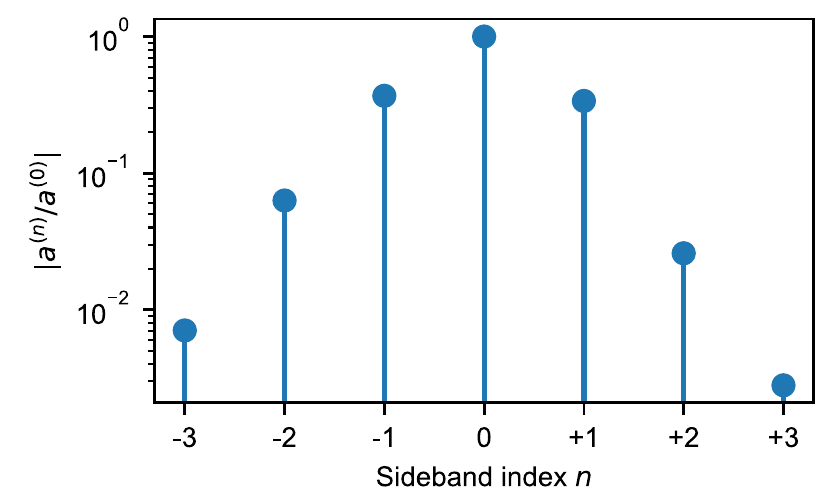}
	\caption{Sideband amplitude distribution in the left resonator of the photonic crystal circulator implementation. The distribution was numerically computed in COMSOL Multiphysics under on-resonance excitation of the compound even mode from the left port.}
	\label{fig:figS3}
\end{figure}

In purely sinusoidal phase modulation, the distribution of the sideband amplitudes is given by the Jacobi-Anger expansion \cite{yariv_quantum_1989}. Fig. \ref{fig:figS3} plots the sideband amplitude distribution inside one of the modulated photonic crystal resonators (from Fig. 4 of the main text) at steady state when excited on resonance. The simulation includes a total of seven frequency components, but the configuration of the system (Fig. 1) allows only the $n=0$ and $n=+1$ bands to couple into and out of the system.

Note that the amplitudes shown in Fig. \ref{fig:figS3} are normalized to the fundamental $n=0$ amplitude.

\section{Modulated Harmonic Oscillator Stability}
\begin{figure}
	\centering
	\includegraphics{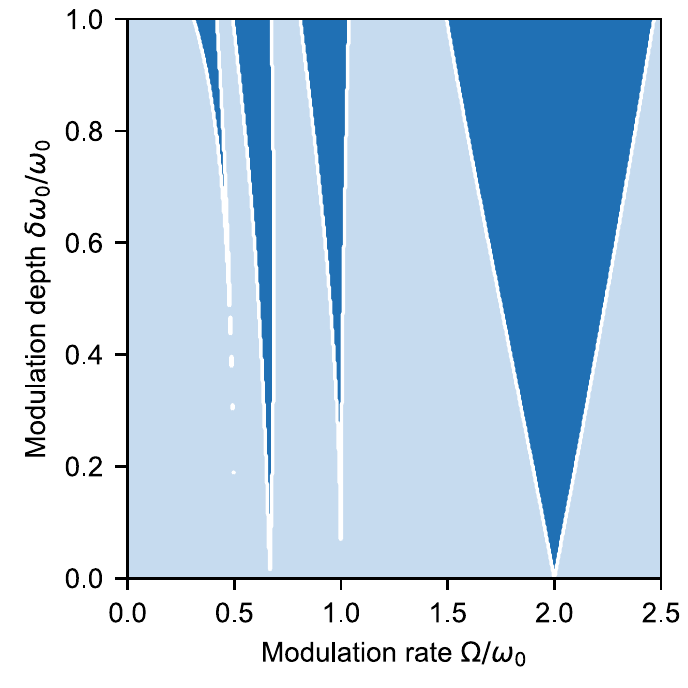}
	\caption{Modulated harmonic oscillator stability map. (light blue) Stable and (dark blue) unstable solutions as a function of modulation rate and modulation depth.}
	\label{fig:figS4}
\end{figure}

Neglecting damping, the Floquet resonator considered in this work is equivalent to a parametrically modulated harmonic oscillator, which obeys the Mathieu differential equation. In dimensionless form this is given by \cite{mclachlan1947theory,acar_floquet-based_2016}
\begin{equation}
\frac{{{\delta }^{2}}a}{\delta {{t}^{2}}}+\left[ 1+\delta \cos \left( \Omega t \right) \right]a=0,
\end{equation}
where $a$ is the oscillator amplitude, $\delta$ is the modulation index, and $\Omega$ is the modulation rate. This equation admits stable and unstable periodic solutions depending on the combination of modulation parameters. A map of the stability regions has been computed and is given in Figure \ref{fig:figS4} where the dark blue regions correspond to unstable solutions that occur due to parametric resonance when the system is driven at harmonics of the fundamental system resonance. In this work we limit consideration to relatively \textit{weak} and \textit{slow} modulation which corresponds to the region around the origin (bottom left) of Figure \ref{fig:figS4}.

\section{Modulated Harmonic Oscillator Solutions}
\begin{figure}[ht!]
	\centering
	\includegraphics{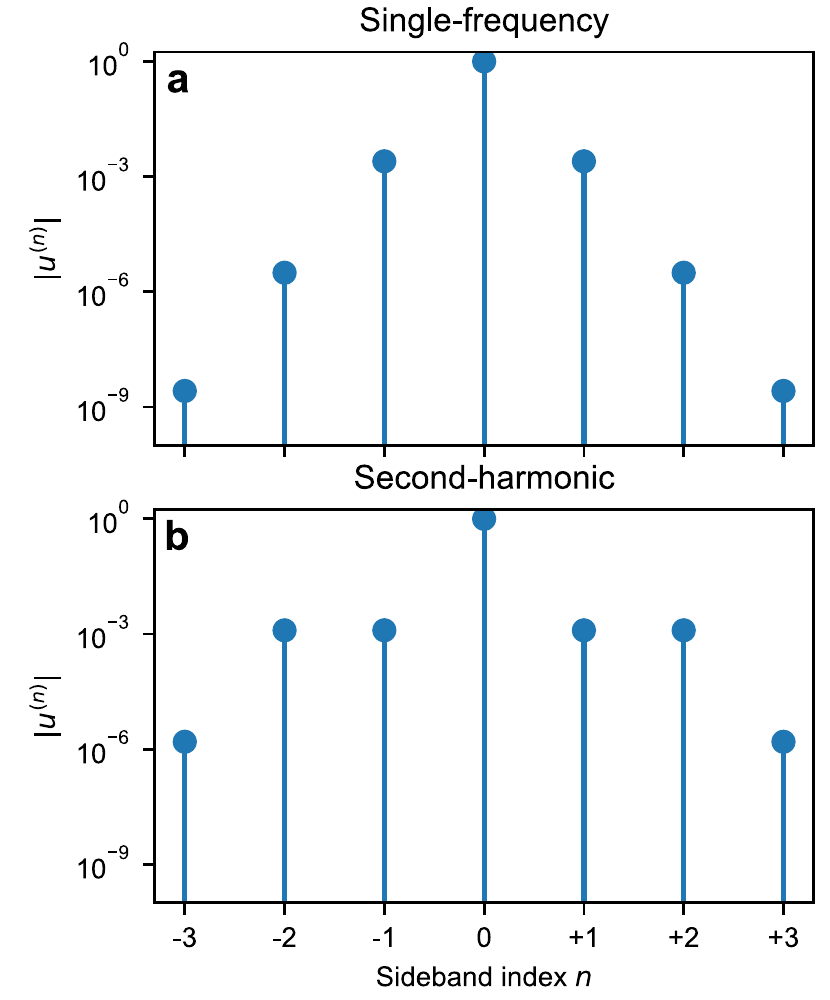}
	\caption{Floquet sideband amplitude distributions for single frequency and second harmonic parametric modulation. Single frequency modulation (left) has modulation of the form $f\left( t \right)=1+\delta \cos \left( \Omega t \right)$ and second harmonic (right) has the form $f\left( t \right)=1+{{\delta }_{1}}\cos \left( \Omega t \right)+{{\delta }_{2}}\cos \left( 2\Omega t \right)$ where $\delta_2=2\delta_1$. Sideband amplitudes were computed by numerically solving Eqn.S31 using the harmonic balance method \cite{mclachlan1947theory,acar_floquet-based_2016}.}
	\label{fig:figS5}
\end{figure}

As discussed in the main text, more general forms of periodic modulation, beyond the case of a sinusoid with a single frequency, can facilitate the control over sideband amplitudes. For example, a modulation waveform that includes a second harmonic component, meaning that it has frequency components of $\Omega$ and $2\Omega$, supports a sideband distribution where the $n=+1$ and $n=+2$ sidebands have the same amplitude, i.e. $u^{\left( 1 \right)} = u^{\left( 2 \right)}$ as shown in Fig. \ref{fig:figS5}.

\section{Generation of Compound Multi-Frequency Mode}
\begin{figure}[ht!]
\centering
\includegraphics{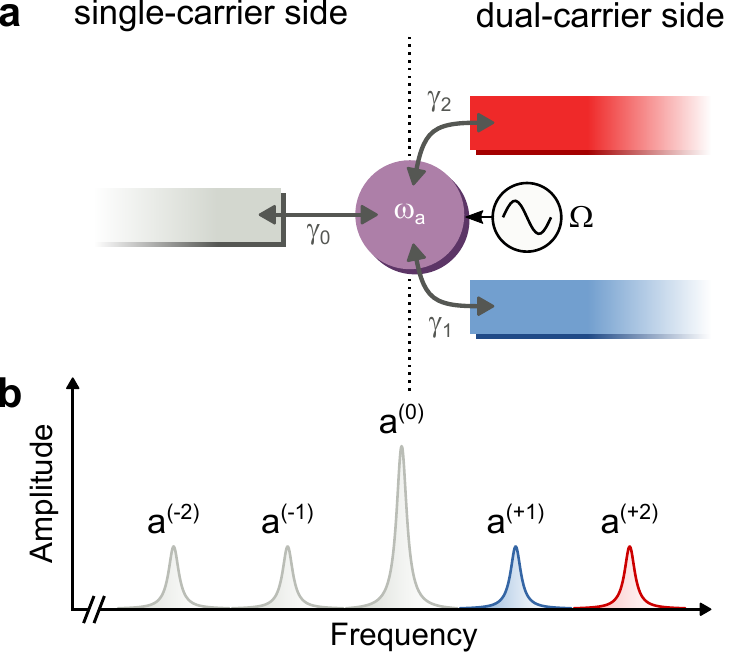}
\caption{Compound multi-frequency mode generation. (a) Schematic of a three-port compound mode generator. (b) Sideband distribution that would be compatible with the structure in (a).}
\label{fig:figS6}
\end{figure}

The compound multi-frequency mode can be generated with the three port optical circuit shown in Figure \ref{fig:figS6}. The signal incident through port one can be resonantly converted into the compound mode defined over ports two and three. This implementation requires that $\gamma_0=\gamma_1+\gamma_2$ which can be achieved through structural and sideband engineering. For example, the Floquet amplitude distribution in Fig. \ref{fig:figS6}b could be used.

Not that this system is reciprocal in that it can also operate in the reverse direction. Incidence of the even mode from the right will be resonantly converted into the single-carrier mode on the left.

\section{Modulation Absorption}
The photonic crystal Floquet circulator discussed in the main text demonstrates the operating principals using full-wave physics but neglects material absorption from modulation. To present a complete picture of the system performance, in this section we characterize the losses that would result from carrier injection and depletion in silicon. This is the strongest mechanism available in silicon, but other approaches and material systems could be the subject of future study.

\begin{figure*}[b!]
	\centering
	\includegraphics{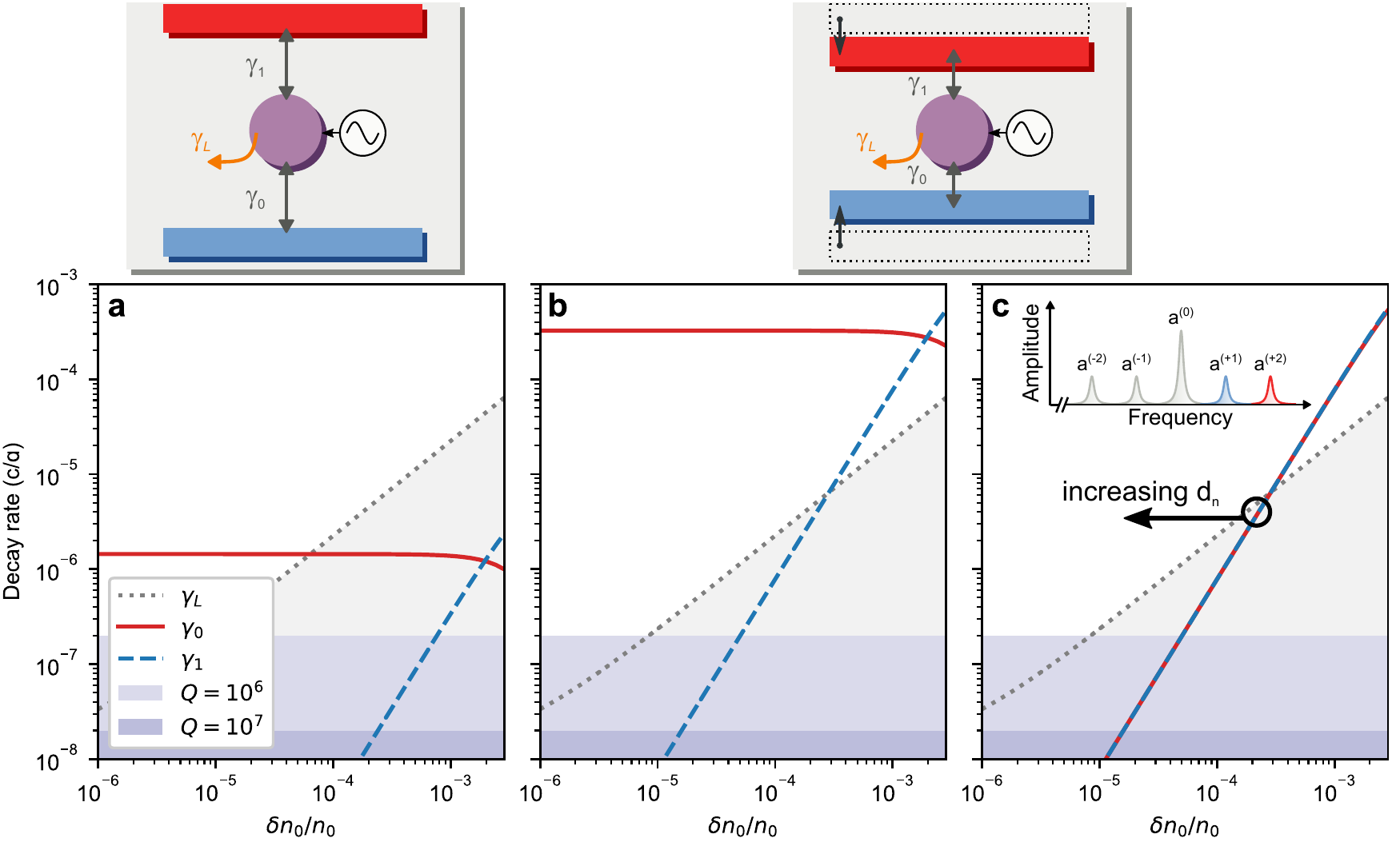}
	\caption{Resonator decay rates to bottom waveguide $\gamma_0$, top waveguide $\gamma_1$, and absorption $\gamma_L$. The top edges of the shaded regions represent the decay rates associated with an out-of-plane quality factor of $10^6$ and $10^7$, demonstrating that the typical values achieved in photonic crystal slab resonators will not overwhelm the circulator response. The absorption rate (dashed border of gray region) is the intrinsic decay rate calculated from an eigenmode simulation of the photonic crystal point defect. A dielectric loss tangent is applied to the simulated defect rod and by assuming modulation takes place through the plasma dispersion effect in silicon, the loss tangent was converted into a relative change in refractive index through the carrier concentrations reported in \cite{soref_electrooptical_1987-1}. (a) Decay rates from fitting to photonic crystal simulation where $d_0=1.2\times10^{-3}$ and $d_1=4\times10^{-3}$. (b) Configuration where $d_0$ and $d_1$ have been increased to $d_0=1.8\times10^{-2}$ and $d_1=6\times10^{-2}$. (c) Configuration where $d_0=1.8\times10^{-2}$ and $d_1=6\times10^{-2}$ and the top and bottom waveguides have been configured to use the first- and second-order Floquet states shown in the inset, where $u^{\left(1\right)} = u^{\left(2\right)}$. In this configuration $\gamma_0$ and $\gamma_1$ are overlapping and critical coupling is achieved for many possible modulation strengths. Further increasing the structural coupling coefficients $d_n$ will shift the curve to the left.}
	\label{fig:figS7}
\end{figure*}

The associated resonator absorption rate is calculated from an eigenmode simulation of the photonic crystal resonator with a dielectric loss tangent applied to the defect rod. The applied loss tangent is converted into a change in refractive index through the carrier concentrations reported in \cite{soref_electrooptical_1987-1}. As shown by the dashed border of gray region in \ref{fig:figS7}a, the resulting absorption rate overwhelms the waveguide coupling rates $\gamma_L \gg \gamma_0,\ \gamma_1$ and means that slight adjustment of the system is needed for practical operation.

One option for overcoming the losses is to increase the structural coupling factors, fitted from the system in Fig. 5 as $d_0=1.2\times10^{-3}$ and $d_1=4\times10^{-3}$. This situation is shown in Fig. \ref{fig:figS7}b where the coefficients have been increased to $d_0=1.8\times10^{-2}$ and $d_1=6\times10^{-2}$ which can be achieved by removing a photonic crystal lattice constant separating the resonator and waveguides. However, this results in a resonator quality factor that's reduced by a factor of $\sim 10^2$, and requires that other aspects of the system be reconfigured. For example, both the modulation frequency and CROW bandwidth would need to be enlarged in order to meet the requirement of one-to-one coupling between sideband states and waveguides.

As discussed in the main text, more general periodic modulation waveforms can be used to provide additional degrees of freedom for tuning the coupling. As shown in Fig. \ref{fig:figS7}c, a sideband distribution with equal amplitudes in $u^{\left(1\right)}$ and $u^{\left(2\right)}$ can be used. By reconfiguring the top and bottom waveguide to couple to $u^{\left(2\right)}$ and $u^{\left(1\right)}$, respectively, the modulation index can be used to tune only the total quality factor. This approach achieves critical coupling between the waveguides and the resonators for a wide range of modulation index because $\gamma_0$ and $\gamma_1$ as a function of modulation index overlap. Increasing the structural coupling coefficients $d_n$ would shift the curve for overlapping $\gamma_0$ and $\gamma_1$ in Fig. \ref{fig:figS7}c further to the left. This means that the circulator can operate with a \textit{total} linewidth comparable to the one in the simulated photonic crystal, a practical modulation frequency, and low loss.

\begin{figure}
	\centering
	\includegraphics{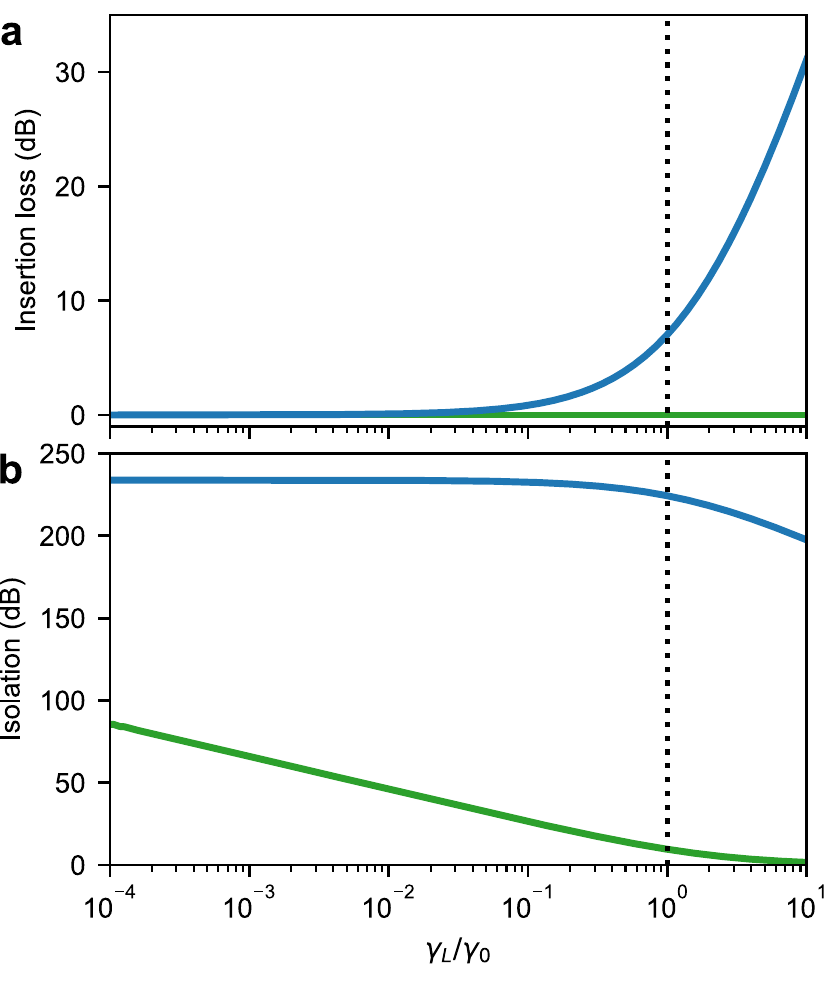}
	\caption{(a) Insertion loss and (b) isolation in the $\alpha_O \to \beta_O$ (blue) and $\beta_E \to \alpha_O$ (green) scattering pathways as a function of the resonator loss rate, normalized to the coupling rate to the bottom waveguide. The $\alpha_O \to \beta_O$ pathway experiences large insertion loss for higher $\gamma_L$ but maintains large isolation. On the other hand, the $\beta_E \to \alpha_O$ pathway experiences reduced isolation for higher $\gamma_L$ but no increased insertion loss.}
	\label{fig:figS8}
\end{figure}

In terms of the isolation and insertion loss performance, the absorption impacts the two cross-device scattering pathways in different ways (Fig. \ref{fig:figS8}). As $\gamma_L$ becomes comparable to $\gamma_0$ and $\gamma_1$, the $\alpha_O \to \beta_O$ pathway experiences higher insertion loss but maintains a very high level of isolation. However, the $\beta_E \to \alpha_O$ pathway experiences reduced isolation for larger $\gamma_L$ with no penalty in insertion loss.

\bibliography{si_refs}
\end{document}